\documentclass{article}
\usepackage[utf8]{inputenc}
\usepackage{pdfpages}

\title{On allocations that give intersecting groups their fair share}
\author{Uriel Feige\thanks{uriel.feige@weizmann.ac.il}~  and Yehonatan Tahan\thanks{yehonatan.tahan@weizmann.ac.il} \\
Weizmann Institute}

\usepackage[numbers]{natbib}
\usepackage{graphicx}
\usepackage{amsmath}
\usepackage{amsthm}
\usepackage{amssymb}
\usepackage{upgreek}
\usepackage{geometry}
\usepackage{multirow}
\usepackage{subcaption}
\usepackage{xcolor}
\usepackage{mdframed}
\usepackage{comment}
\usepackage{diagbox}
\usepackage{float}
\usepackage{xfrac}
\usepackage{dsfont}
\usepackage{mathtools}

\usepackage{hyperref}

\newcommand\TODO[1]{\textcolor{red}{TODO: }(#1)}
\newcommand\explanation[1]{& \color{gray}{#1}}
\newcommand{\brs}[1]{\left[ {#1} \right]}
\newcommand{\prs}[1]{\left( {#1} \right)}
\newcommand{\abs}[1]{\left| {#1} \right|}
\newcommand{\curprs}[1]{\left\{ {#1} \right\}}
\DeclarePairedDelimiter\ceil{\lceil}{\rceil}
\DeclarePairedDelimiter\floor{\lfloor}{\rfloor}
\def\curlH{\mathcal{H}}
\def\curlG{\mathcal{G}}
\def\curlA{\mathcal{A}}

\def\NN{\mathbb{N}}

\newcommand{\items}{{\cal{M}}}
\newcommand{\agents}{{\cal{N}}}
\newcommand{\groups}{{\cal{G}}}
\newcommand{\groupsNum}{g}

\newcommand{\uniform}{{uniform }}
\newcommand{\Uniform}{{Uniform }}
\newcommand{\unif}{{uniform}}

\newtheorem{theorem}{Theorem}[subsection]
\theoremstyle{definition}
\newtheorem{definition}{Definition}[subsection]
\newtheorem{corollary}{Corollary}[subsection]
\newtheorem{lemma}{Lemma}[subsection]
\newtheorem{claim}{Claim}[subsection]
\newtheorem{observation}{Observation}[subsection]

\newtheorem{proposition}{Proposition}[subsection]
\newtheorem{fact}{Fact}[subsection]
\newtheorem{example}{Example}[subsection]
\newtheorem*{remark}{Remark}
\newtheorem*{fact*}{Fact}

\newtheorem*{theorem*}{Theorem}

\usepackage[]{color-edits}
\addauthor{UF}{red}

\addauthor{YT}{blue}

\newcommand{\yte}[1]{{\YTedit{#1}}}
\newcommand{\specialcell}[2][c]{%
  \begin{tabular}[#1]{@{}c@{}}#2\end{tabular}}

\begin{document}

\maketitle

\begin{abstract}
{We consider item allocation to individual agents who have additive valuations, in settings in which there are protected groups, and the allocation needs to give each protected group its
``fair" share of the total welfare.}
Informally, within each protected group we consider the total welfare that the allocation gives the members of the group, and compare it to the maximum possible welfare that an allocation can give to the group members. An allocation is fair towards the group if the ratio between these two values is no worse then the relative size of the group. For divisible items, our formal definition of fairness is based on the proportional share, whereas for indivisible items, it is based on the anyprice share.  

We present examples in which there are no fair allocations, and even not allocations that approximate the fairness requirement within a constant multiplicative factor. We then attempt to identify sufficient conditions for fair or approximately fair allocations to exist. 
For example, for indivisible items, when agents have identical valuations and the family of protected groups is laminar, we show that if the items are chores, then an allocation that satisfies every fairness requirement within a multiplicative factor no worse than two exists and can be found efficiently, whereas if the items are goods, no constant approximation can be guaranteed.
\end{abstract}

\section{Introduction}


The task of allocating resources to individual parties takes a major role in our lives, from budget distribution to divorce settlements, from land division in peace treaties to course allocation in universities. In some of such scenarios the resources are divisible (money, land, time on a computing resource) and some involve resources that cannot or should not be broken to pieces  (items of sentimental value in inheritance, vacant slots in courses). 
The question of how to allocate resources in a manner that is fair towards individual agents has been studied extensively in the past few decades, both for divisible and indivisible resources. See for example~\cite{Dubins_Spanier_divisible, Lipton_envy_cycle_indivisible, Budish_MMS_EF1}.
Some more recent work(~\cite{WMMS_def, BEF21, WEF1_def}) considered the case of allocating the resources to participants that might have different entitlements.

We consider a setting of fair allocation of items to agents, when there are protected groups of agents, and the allocation needs to satisfy certain fairness requirements towards the protected groups. Before introducing our setting, let us briefly recall a well studied setting for item allocation in which there are no groups. There is a set $\items$ of $m$ items, and a set $\agents$ of $n$ agents. Every agent $i$ has an additive valuation function $v_i$ over the set of items. Namely, $v_i(S) = \sum_{e \in S} v_i(e)$, for every $S \subset \items$. For divisible items, the value of an $\alpha$-fraction of an item is an $\alpha$-fraction of the full value of an item. An {\em allocation} $A = (A_1, \ldots, A_n)$ is a partition of $\items$ into $n$ disjoint bundles, where agent $i$ receives bundle $A_i$. A {\em fractional allocation} allows fractions of items to be allocated to the agents, where for every item, the sum of allocated fractions is~1. For simplicity of the presentation at this stage, assume that items are {\em goods} ($v_i$ is non-negative), though at a later stage we will also consider {\em chores} ($v_i$ is non-positive). As our setting does not involve transfer of money, we wish our allocation to meet some fairness benchmark. This benchmark is expressed in form of a {\em share}, where a share function $s$ maps that valuation function of the agent and her entitlement (for simplicity, assume that all agents are equally entitled, and then the entitlement is $\frac{1}{n}$) to a real value, and the intended interpretation of $s(v_i, \frac{1}{n})$ is that agent $i$ is untitled to receive a bundle $A_i$ satisfying $v_i(A_i) \ge s_i(v_i, \frac{1}{n})$. An allocation that gives every agent at least her share is {\em admissible}, and for every $0 < \rho \le 1$, an allocation that gives every agent at least a $\rho$ fraction of her share is $\rho$-admissible. Common examples of shares are the proportional share ($s_i(v_i, \frac{1}{n}) = \frac{1}{n} v_i(\items)$) for divisible items, and the maximin share~\cite{Budish_MMS_EF1} (MMS, see definition{~\ref{def:mms}}) for indivisible items. With respect to the above notions of shares, admissible allocations exist in the divisible case (every agent gets a $\frac{1}{n}$ fraction of every item), and $\rho$-MMS allocations exist in the indivisible case, for $\rho = \frac{3}{4}+\frac{1}{12n}${~\cite{0.75_mms_allocation}}.

In our work, the above setting is augmented by a list $\groups = G_1, G_2, \ldots$ of protected groups of agents, where for every $G\in \groups$ we have $G \subseteq \agents$. The requirement is that the allocation be fair towards each of the protected groups. Our notion of fairness towards groups is share based, similar to our fairness notion towards individual agents. For this purpose we extend valuation functions to groups. Our extension aims to capture the intention that within a group items should be allocated efficiently, namely, to those agents in the group that value the items more. Hence we define the value of item $e$ for group $G$ to be $v_G(e) = \max_{i\in G} v_i(e)$, and this extends in an additive (linear) way to values of sets of items and/or fractions of items. We also need to define the entitlement of a group, and we define it to be its relative size ($\frac{|G|}{n}$, for group $G$). Given the above definitions, we consider the proportional share (PS) when items are indivisible, and the anyprice share~\cite{BEF21} (APS, see definition{~\ref{def:aps for groups}}) when items are indivisible. An allocation is admissible if for every group, the sum of values received by its members is at least as high as the share of the group, under the above definition. For example, for divisible items, an allocation $A = (A_1, \ldots, A_n)$ is admissible if for every group $G \in \groups$ it holds that $\sum_{i\in G} v_i(A_i) \ge \frac{|G|}{n} v_G(\items)$, where $v_G(\items) = \sum_{e\in \items} \max_{i\in \agents} v_i(e)$. The definition of admissible allocations is extended to $\rho$-admissible ones, in an analogous way to that presented above.

Two comments are in order here. One is that admissible allocations are required to give the protected groups at least their share, but there is no requirement towards non-protected groups. In particular, an individual agent is required to get at least her share only if the agent is listed as a protected group (we allow groups of size~1). The other is that the definition of the group valuation involves summing up valuations of different group members. For such a sum to make sense in those cases in which different agents have different valuation functions, we need to assume that all valuation functions are measured in the same units, such as units of money in a given currency.

We now present a sequence of examples for our setting, where each new example adds one level of complication over the preceding one.

\begin{example}
\label{ex:disjoint}
{\bf Disjoint protected groups.} Several parties in the Parliament join to form a coalition government. There are certain ``luxurious jobs" (which we refer to as {\em items}) that are to be assigned to individual Parliament members from the coalition parties (we refer to the eligible individuals as {\em agents}), such as speaker of the house, heads of various parliamentary committees, etc.  Each agent may have individual preference over items. The allocation of items is required to be fair towards the parties in the coalition: the overall quality of items that the members of a party receive (where quality is measured in the eyes of members of the coalition party) should be proportional to its size. The protected groups in this example are the parties of the coalition.
   
In the above example, the protected groups are disjoint. The case of disjoint protected groups reduces to the setting of item allocations to agents with arbitrary (unequal) entitlement (without augmenting this setting with protected groups), where each protected group can be thought of as an individual agent with entitlement proportional to the size of the group, and having a valuation function under which the value of a set of items is the maximum welfare the group can receive by allocating the items to those members of the group that value them most. 
\end{example}

\begin{example}
\label{ex:conflict}
{\bf Two level hierarchy.} Extending the setting of Example~\ref{ex:disjoint}, suppose that within each party in the coalition, not all members have equal entitlement. For example, the head of the party may be viewed as having higher entitlement to a ``luxurious job" then other members of the party. We may model this by dividing the total entitlement of the party (its relative size) in an unequal way among its members. Moreover, now every individual agent (or at least some individual agents, such as heads of parties) becomes a protected group, in the sense that the item allocation need not only  be fair towards each party as a whole, but also towards (all or some) individual agents within the parties. Now the protected groups form a hierarchy with two levels, individual agents at the bottom level, and coalition parties at the top level. Having two levels may create conflicts for the allocation that are not present when there is only one level. The allocation of items within a party needs to satisfy two possibly conflicting goals. One is to allocate them to those members who value them most, so that the party as a whole gets high welfare. The other is to allocate them to those agents that have higher entitlement, so that these agents get their fair share. A conflict might exist even if all agents within a protected group have the same valuation function. For example, consider a protected group $G$ composed of two protected agents with identical valuations. Given the option to give $G$ either one item of value~3 or two items of value~1, the first option is preferable in terms of fairness towards $G$, whereas the second option may be preferable in terms of fairness towards the protected agents (as then each of the agents can get one item). 

Another (more standard) example of a two level hierarchy is a setting of allocating items to individual agents (with no protected groups) in which there are two requirements: one is to be fair towards the individual agents (giving each agent at least her share, or at least approximately her share), and the other is to maximize welfare (or approximately maximize welfare). In our terminology, this is simply the setting in which the protected groups form a hierarchy with two levels, where protected groups at the bottom layer are the individual agents, and there is one protected group at the top level, which is the set of all agents.
\end{example}

\begin{example}
\label{ex:hierarchy}
{\bf Multi-level hierarchy.} Extending the setting of Example~\ref{ex:conflict}, suppose that some of the coalition parties are composed of smaller political parties that were merged prior to the elections. Within such a coalition party, it may be required that item allocation respects the relative sizes of the smaller political parties from which it is composed. So now we have three levels of protected groups: the coalition parties, the smaller political parties that make up each coalition party, and the individual members of the small political parties. Having a multi-level hierarchy of protected groups might amplify the conflicts referred to in Example~\ref{ex:conflict}. 

Multi-level hierarchies may come up naturally in other settings, such as in allocating resources (such as supplying NMR machines) to medical centers, where one wants to be fair towards individual medical centers (those that serve more patients have higher entitlement), fair towards cities (more populated cities have higher entitlement), fair towards counties, and fair towards states. 
\end{example}

\begin{example}
\label{ex:general}
{\bf Arbitrary intersection pattern.} Extending the setting of Example~\ref{ex:hierarchy}, suppose  that there are protected groups that transcend party lines, such as the group of women in the coalition parties, or the group of agents from some underprivileged minority group. In such settings, the protected groups do not form a hierarchy, and may have arbitrary intersection patterns.
\end{example}

As mentioned in Example~\ref{ex:conflict}, when protected groups intersect, this raises conflicts as to how to allocate items in this intersection. This conflict might exist even if the valuation of all agents in the intersection are identical, and may become more severe the larger the differences among valuation functions. In our work we consider several families of valuation functions, capturing various levels of similarities among valuation functions, and similarities within valuation functions.

\begin{itemize}
    \item {\bf \Uniform valuations.} $v_i(e)=1$ for every agent $i$ and item $e$. In this setting the fairness requirement depends only on how many items each protected group receives.
    \item {\bf Dichotomous valuations.} For every agent $i$ and item $e$, either $v_i(e)=1$ or $v_i(e)=0$. Also in this setting the fairness requirement depends only on how many items each protected group receives, but now we only count those items that have value~1 for at least one group member. For previous work on dichotomous valuation functions, {see for example~\cite{BEF21a}}.
    \item {\bf Identical valuations.} There is some additive valuation function $v$, and all agents have the same valuation function. In this setting, for every protected group, the fairness requirement only concerns which items the group receives, but not how these items are allocated to members within the group.
    \item {\bf Restricted assignment.} There is some additive valuation function $v$, and for every agent $i$ and item $e$,  either $v_i(e) = v(e)$ or $v_i(e)= 0$. This generalizes both dichotomous valuations (in which $v(e) = 1$ for every item) and identical valuations (in which $v_i(e) = v(e)$ for every agent $i$ and item $e$). For previous work on restricted assignment valuation functions, see for example~\cite{BS06}.
    \item {\bf Scaled valuations.} For every two agents $i$ and $j$ it holds that $v_i(\items) = v_j(\items)$. As remarked above, our definition of valuation functions for the protect groups makes sense only if valuation functions of different agents are measured in the same units. For scaled valuations, these units are ``fractions of value of the grand bundle". 
    In particular, \uniform valuations are scaled.
\end{itemize}


Given a set $\items$ of items and a set $\agents$ of agents, one may ask how one determines which groups are protected. We envision two conceptually different procedures for doing so (though of course, a combination of the two is also possible). In an internal procedure for determining the protected groups, the agents themselves are those that provide the list of protected groups. For example, every agent may be allowed to list one group of her choice, and this group becomes protected. In Example~\ref{ex:hierarchy}, it could be that the reason that a small political party becomes protected is because one of its members listed it as protected. If no agent lists this small political party as protected, then it is not protected. If an agent $i$ lists a group other than herself, to the other agents this may appear to be a display of {\em altruism} (regardless of whether agent $i$'s underlying motivation for listing the group is altruistic or egoistic), as the agent herself might then remain unprotected, and only the group that she (supposedly) cares about is protected. In an external procedure for determining the protected groups, the composition of the protected groups is determined externally, and does not depend on preferences of the agents. In Example~\ref{ex:general}, it could be that women are a protected group due to some regulation, and the allocation needs to respect this, even if the agents themselves are indifferent towards (or even disagree with) this regulation.



In this paper we investigate fair item allocation questions when there are protected groups. For concreteness, we assume the internal procedure for determining the protected groups, in which every agent $i$ reports a group $G_i$ of agents that $i$ wishes to protect.  This group may be $i$ itself ($G_i = \curprs{i}$), in which case we may think of $i$ as being selfish, may be some larger group that includes $i$, or may even be a group that does not include $i$ (a report that appears to be {\em altruistic}). {Note that under the internal procedure for determining the protected groups, the number $\groupsNum$ of protected groups is at most $n$. Most of our results hold as stated also $\groupsNum > n$, and the others can be extended with minor modifications to this case, though details of these extensions are sometimes omitted.}
We study both existential and computabilty questions of allocations that are fair towards all groups simultaneously. In order to define what is a fair allocation, we distinguish between divisible and indivisible items. For divisible items we take the \emph{proportional share} as a fairness objective, and for indivisible items we adapt the notion of \emph{anyprice share}~\cite{BEF21}. 
In more detail, we are interested in what is the best fraction of the \emph{anyprice share} (\emph{proportional share}) that can be guaranteed to every voted group by an integral (fractional) allocation of the indivisible (divisible) items. In addition, we discuss the case of items that are \emph{chores},  and we also discuss the possible manipulation of allocation mechanisms (that accommodate reporting of groups) by strategic agents who misreport their true protected groups.

\subsection{Definitions}

{An Item allocation instance consists of a set ${\items}=\curprs{e_1,e_2,\dots, e_m}$ items that should be allocated to a set ${\agents}=\curprs{1,2,\dots,n}$ of agents. Each agent $i$ has a valuation function $v_i:2^{\items} \rightarrow {\mathbb{R}}_{\geq 0 }$. Throughout this work, we assume that all valuations are normalized ($v_i \prs{\emptyset} = 0$) and additive ($v_i \prs{B} = \sum_{e\in B} v_i \prs{\curprs{e}}$). For ease of notation, we will use $v_{i,e}$ and $v_i \prs{e}$ instead of $v_i \prs{\curprs{e}}$.
An integral allocation $A$ is a vector of $n$ disjoint bundles $\prs{A_1,A_2,\dots,A_n}$ where $A_i\subseteq {\items}$. $A_i$ is the bundle of items allocated by $A$ to agent $i$. The welfare agent $i$ gains from an integral allocation $A$, denoted as $w_i \prs{A}$ is $v_i \prs{A_i}$.
A fractional allocation, can be represented as a non-negative matrix $X \in [0,1]^{{\agents \times \items}}$ such that $\forall e\in {\items}: \sum_{i\in {\agents}} X_{i,e} \leq 1$. The value $X_{i, e}$ is the fraction of the item $e$ that is allocated to agent $i$ by $X$. Notice that an integral allocation can be represented as a fractional one, using $X_{i,e} = 1 \iff e\in A_i$. The welfare agent $i$ receives from a fractional allocation $X$, denoted as $w_i \prs{X}$, is $\sum_{e\in {\items}} {X_{i,e} \cdot v_i\prs{e}}$.
When items are indivisible, we consider only integral allocations. When dealing with divisible items, we consider fractional allocations as well.
{We consider settings with no monetary transfers (neither between agents nor payments).  In such settings, it is natural to seek allocations that are both Pareto optimal (no other allocation Pareto dominates it, i.e. in every other allocation either at least one agent is strictly harmed by the change of allocations or no agent  strictly benefits from this change), and provide some measure of fairness.}


{Before turning to define protected groups in such a setting and appropriate notions of fairness there, let us briefly review some fairness notions that have been studied for item allocation instances. {There are two main types of fairness notions that have been studied extensively, those that are \emph{envy} based (attempt to avoid envy between participants) and \emph{share} based (attempt to give participants their fair-share). We start with the \emph{envy} based ones.} An integral allocation $A$ is \emph{envy-free} (EF) if $\forall i, j\in {\agents}: v_i\prs{A_i}\geq v_i\prs{A_j}$. A similar definition holds for fractional allocations as well. 
As EF can be unattainable via integral allocation, a relaxed notion of \emph{envy-free up to one good} (EF1) was proposed by Budish~\cite{Budish_MMS_EF1} -- an allocation $A$ is EF1, if any envy can be alleviated by removing at most one item from the bundle received by the envied agent. EF1 integral allocations always exist~\cite{ef1_exists}.

We now discuss \emph{share} based fairness notions. The \emph{proportional share} (PS) of agent $i$ is defined to be $PS\prs{i}=\sfrac{v_i\prs{{\items}}}{n}$. When items are divisible, the entire PS can be guaranteed by giving every agent a $\sfrac{1}{n}$ fraction from every item. For indivisible items, the entire PS cannot always be guaranteed (consider two agents and a single item that both agents value as 1), and neither can any positive fraction of it. 
To handle indivisible items, a different share notion was proposed by Budish~\cite{Budish_MMS_EF1}: 

\begin{definition}[Maximin Share]
\label{def:mms}
The \emph{maximin share} of an agent $i$ in an instance with $n$ agents is defined to be
\begin{equation*}
    MMS_i = \max_{A = \prs{A_1,\dots,A_n}}{\min_{j \in {[n]}} v_i\prs{A_j}}
\end{equation*}
    
\end{definition}

An allocation $A$ is $MMS$-fair if $\forall i\in {\agents}: w_i\prs{A}\geq MMS_i$. An allocation $\rho$-approximates the MMS if $\forall i\in {\agents}: w_i\prs{A}\geq \rho \cdot MMS_i$. For agents with additive valuations, there are instances in which there is no $MMS$-fair integral allocation~\cite{no_full_MMS},  but there always is an integral allocation giving every agent a constant fraction of her MMS. The best $MMS$-approximation guarantee currently known is $\frac{3}{4} + \Omega\prs{\frac{1}{n}}$ (\cite{0.75_mms_allocation}). 

The MMS and EF1 are incomparable fairness notions. 
An MMS allocation need not to be EF1, and an EF1 allocation might give an agent only a $O(\sfrac{1}{n})$ fraction of her MMS, even if the allocation is Pareto optimal. In our work, we consider only share based fairness notions.
}

Let us now define what is an item allocation instance that involves {protected groups}.


\begin{definition}[Item Allocation with Protected Groups]
An \emph{item allocation instance with protected groups} is an item allocation instance, associated with a \emph{reference family} $\curlG = \curprs{G_1,G_2,\dots,G_{{\groupsNum}}}$ where $G_i \subseteq {\agents}$ is {a protected group}. 
\end{definition}

The welfare group $G\subseteq {\agents}$ receives from specific (fractional or integral) allocation $X$ is $w_{G}\prs{X} = \sum_{i\in G} w_i\prs{X}$.
In order to assess a particular allocation in the eyes of the reference groups, we need to establish first a valuation function. We propose to define the group valuation for a bundle $B$, as the optimal welfare it can gain from $B$. Formally, for a group $G\subset {\agents}$ and an item $e\in {\items}$ we define $v_{G}{\prs{e}} = \max_{j\in G} v_{j}\prs{e}$. 

Using this definition of group valuation, we turn to define fairness notions for groups of agents. We adapt notions suitable to item allocation settings with agents that have unequal entitlement.

\begin{definition}[group proportional share]
Given an instance of item allocation $\prs{\agents, \items, v_1,\dots,v_n}$ and a group $G\subseteq \agents$, the \emph{proportional share} of $G$ is $PS\prs{G} = \frac{\abs{G}}{n}\cdot v_G\prs{\items}$.
\end{definition}

This notion is useful when dealing with divisible items. Given an \emph{item allocation with {protected groups}} instance with divisible items, we say that a fractional allocation $X$ $\rho$-approximates the \emph{proportional share} if $\forall G\in \curlG: w_G\prs{X} \geq \rho \cdot PS\prs{G}$. As mentioned in Proposition~\ref{prop: Divisible LP}, given a specific instance, we can efficiently compute a fractional allocation the maximizes $\rho$.

For indivisible items, we adapt the \emph{anyprice share} notion, introduced by Babaioff, Ezra and Feige(~\cite{BEF21}) for item-allocation instance of indivisible items with agents that have arbitrary entitlement. We use  $\frac{\abs{G}}{n}$ as the group entitlement.

\begin{definition}[group anyprice share]
\label{def:aps for groups}
Given an instance of item allocation $\prs{\agents, \items, v_1,\dots,v_n}$. Denote ${\mathcal{P} = \curprs{\prs{p_1,p_2,\dots,p_m}: \forall i, p_i \geq 0, \sum_{i=1}^{m}{p_i} = 1}}$. The \emph{anyprice share} of $G$ is defined as: 
\begin{equation*}
    APS\prs{G} = \min_{\prs{p_1,\dots,p_m}\in \mathcal{P}} \max_{B\subseteq \items}\curprs{v_G\prs{B}: \sum_{e_t\in B}p_t \leq \frac{\abs{G}}{n}}
\end{equation*}
\end{definition}

We make use of this definition when the instance at hand involves indivisible items. Given an \emph{item allocation with {protected groups}} instance with indivisible items, we say that an integral allocation $A$ $\rho$-approximates the \emph{anyprice share} if $\forall G\in \curlG: w_G\prs{X} \geq \rho \cdot APS\prs{G}$.

We remark that unlike divisible items, given a specific instance $\prs{\agents,\items,\groups, v_1,\dots,v_n}$, computing the integral allocation that yields the maximal $\rho$-approximation to the $APS$ is $NP$-hard, even if $\curlG$ forms a partition of $N$.

In this work, we study, for various types of instances, the question of what maximal $\rho$-approximation can be guaranteed to any reference group in every instance. When ``good" $\rho$-approximation is guaranteed to exist (e.g. $\rho = \Theta\prs{1}$), we also concern ourselves with the question of efficiently computing such an allocation. As in the general case it is NP-hard to compute the best allocation with indivisible items, the latter concern is of most interest when items are indivisible and we are assured that good $\rho $ approximation ratio exists. In these cases, we wish to answer whether finding an allocation that  guarantee such $\rho$-approximation can be done efficiently or not. 

Next we turn to define some restrictions on the reference families and the valuation function.
First, we describe restrictions to the structure of the reference family. If the reference groups are disjoint, then we can convert the instance into a new instance of item allocation to agents with different entitlements. In the new instance every agent $i_G$ represents the group $G$ in the original instance, his valuation function is $v_{i_G} = v_G$ and his entitlement is $\sfrac{\abs{G}}{n}$. From previously known results, we can produce efficiently a fractional allocation that gives each agent $i_G$ (and therefore each group $G$) its entire PS, and an integral allocation that $\sfrac{3}{5}$-approximates the APS. Hence, our work will focus on families of groups that do intersect.
We call a reference family $\curlG \subseteq 2^\agents$ \emph{laminar}, if $\forall G,H\in \curlG$ either $G\cap H = \emptyset$, or $G\subseteq H$, or $H\subseteq G$. This restriction makes sense when $\agents$ has a natural sequence of partitions, where each one refines the next. For example, a company will usually have some levels of partition, e.g. the entire company, departments, groups, teams and finally individual employees. {When the groups are reported by the agents themselves}, one might expect any individual employee to report (as a reference group) either herself, her team, her group, etc. 
Notice that typically the expansion between the levels of the hierarchy will be substantial. The size of the group will probably be much larger than any of its teams. We formally define this phenomena using the notion of $\alpha$-hierarchical families.  A reference family $\curlG \subseteq 2^\agents$ is called \emph{$\alpha$-hierarchical family} for some $\alpha\geq 1$ if it is laminar and $\forall X,Y\in \curlG$ if $X\subsetneq Y$ then $\alpha \cdot \abs{X} \leq \abs{Y}$. We mainly focus on the case where $\alpha = 2$.

When coming to restrict the types of valuation functions, natural restrictions may come from the fact that agents might value items in different currencies. Some of the restrictions that we introduce aim to address this issue. One such restriction ({\em scaled valuations}) is that all agents have the same value to the entire bundle of items, and therefore valuations differ only in terms of the relative value they give to each item. Another such restriction ({\em restricted assignment}) assumes that there is some underlying valuation function {$v$}, and the valuation of each agent is restricted in terms of the way by which it differs from {$v$}. 
In the next definition we introduce several types of restrictions.
}


{
\begin{definition}
\label{def:valTypes}
Let $\mathcal{I} = \prs{\agents, \items, v_1,v_2,\dots, v_n}$ be an item allocation instance.
\begin{itemize}

    \item \textbf{{\Uniform Valuations.}} A valuation function $v$ is \uniform if $v\prs{e}=1$ for every $e \in \items$. The instance $\mathcal{I}$ has \emph{\uniform valuations} if $\forall i\in \agents$ the valuation function $v_i$ is \emph{\unif}.
    \item \textbf{Dichotomous Valuations.} We say that $\mathcal{I}$ has \emph{dichotomous valuations} if  $\forall i \in \agents, {e} \in \items: v_i\prs{j} \in \curprs{0, 1}$.
    \item \textbf{Identical Valuations.} $\mathcal{I}$ has \emph{identical valuations} if $\forall i,j\in \agents: v_i = v_j$.
     \item \textbf{Restricted Assignment.} The instance $\mathcal{I}$ is called a  \emph{restricted assignment} instance if there exist ${v}:\items\rightarrow {\mathbb{R}}_{\geq 0 }$ such that $\forall i\in \agents, {e}\in \items: v_i\prs{{e}}\in \curprs{v\prs{{e}}, 0}$.
     \item \textbf{Scaled Valuations.} We say that $\mathcal{I}$ has \emph{scaled valuations} if $\forall i,j \in \agents: v_i\prs{\items} = v_j\prs{\items}$. {\Uniform valuations are scaled.}

\end{itemize}
\end{definition}
}

{
\subsubsection*{Truthfulness}


When thinking of our model as $n$ players communicating with a designer, a natural question that arises is whether or not a player can abuse the system and gain additional value by not reporting her honest answers to queries from the designer. {In the internal procedure for determining the protected groups}, each player $i$ has two pieces of information associated with her, namely $v_i$ and $G_i$. {Consequently,} player $i$ might have two different ways to manipulate the system, either by misreporting $v_i$, or by misreporting $G_i$.
Here {we explore a setting in which}
the valuations are publicly known as well as $\agents$ and $\items$, and the only private information that each player $i$ is required to report is her reference group $G_i$. 
We now define the notion of a truthful allocation algorithm with respect to reporting of groups.

\begin{definition}[Truthful allocation algorithm with respect to reporting of groups]
An algorithm $\curlA$ operating on \emph{item allocation with {protected groups}} instances is called \emph{truthful with respect to group reporting}, if for every instance $\mathcal{I} = \prs{\agents, \items, v_1,v_2,\dots, v_n, G_1,\dots,G_i,\dots, G_\groupsNum}$, every player $i$ and every group $G'_i$ it holds that:  
\begin{equation*}
    w_{G_i}\prs{\curlA\prs{\mathcal{I}}} \geq w_{G_i}\prs{\curlA\prs{\agents, \items, v_1,v_2,\dots, v_n, G_1,\dots,G'_i,\dots, G_\groupsNum}}
\end{equation*}

\end{definition}

In Section~\ref{subsection: truthfulness results} we discuss how well such truthful algorithms can approximate our share notions (\emph{proportional share} and \emph{anyprice share}).
}

\subsubsection*{Chores}
In Section~\ref{subsection: chores results} we discuss what happens when items are ``bads" or \emph{chores}. This setting is also an item allocation setting with non-negative valuations. However, unlike the setting of allocation of goods, we view the valuations as costs. The larger the value an agent $i$ assigns to chore $e$, the least $i$ prefers to get this chore in the allocation. We present natural adaptations of our group valuation and fairness notions (that we have defined for goods) to the setting of chores. Given an instance $\mathcal{I} = \prs{\agents, \items, v_1,v_2,\dots, v_n}$, where items are chores, and a group of agents $G\subseteq \agents$, $G$'s valuation for a chore $e\in \items$ is $v_G\prs{e} = \min_{i\in G} v_i\prs{e}$. The social cost of $G$ from an allocation $X$ is denoted as $c_G\prs{X}$ and defined as the social welfare $G$ received from $X$. The \emph{proportional share} of $G$ is defined identically to the ``goods" case. The \emph{anyprice share} is defined as follows.

\begin{definition}[group anyprice share for chores]
Given an instance of chores allocation $\prs{\agents, \items, v_1,\dots,v_n}$. Denote ${\mathcal{P} = \curprs{\prs{p_1,p_2,\dots,p_m}: \forall i, p_i \geq 0, \sum_{i=1}^{m}{p_i} = 1}}$. The \emph{anyprice share} of $G$ is defined as: 
\begin{equation*}
    APS\prs{G} = \max_{\prs{p_1,\dots,p_m}\in \mathcal{P}} \min_{B\subseteq \items}\curprs{v_G\prs{B}: \sum_{e_t\in B}p_t \geq \frac{\abs{G}}{n}}
\end{equation*}
\end{definition}

Let  $\mathcal{I} = \prs{\agents, \items, v_1,v_2,\dots, v_n, \groups}$ be an instance of \emph{chores allocation with {protected groups}}. We say that an integral allocation $A$ $\alpha$-approximates the \emph{anyprice share} if $\forall G\in \groups: c_G\prs{X} \leq \alpha \cdot APS\prs{G}$. Similarly, a fractional allocation $X$ $\alpha$-approximates the \emph{proportional share} if $\forall G\in \curlG: c_G\prs{X} \leq \alpha \cdot PS\prs{G}$.

\subsection{Our results}

We divide our main results into two categories: results that address the best approximation ratio (to the \emph{proportional share}) that can be guaranteed in instances with divisible items (in Section~\ref{subsection: divisible items}), and results regarding the best approximation ratio (to the \emph{anyprice share}) that can be guaranteed in instances with indivisible items (in Section~\ref{subsection: indivisible items}). In each of these two settings, we consider the sub-settings of possible restriction combinations of the reference family (general, laminar, 2-hierarchical) and the valuation functions (additive, \unif, dichotomous, identical, restricted assignments, scaled).

Before presenting our results regarding divisible items, we observe that when all valuations are identical the entire \emph{proportional share} can be guaranteed to all groups

\begin{observation}
When items are divisible and all valuation functions are identical ($\exists v:\items\rightarrow {\mathbb{R}}_{\geq 0} \text{ s.t. } \forall i\in \agents: v_i \equiv v$) then the uniform distribution fractional allocation $X$ that is defined by $\forall i\in \agents, e\in \items: X_{i,e} = \frac{1}{n}$ guarantees the entire $PS\prs{G}$ for all $G\subseteq \agents$.
\end{observation}

Following this observation, in Section~\ref{subsection: divisible items} we focus only on divisible-items instances in which the valuation functions are not all identical. {For such instances, the allocation in the above observation ($\forall i\in \agents, e\in \items: X_{i,e} = \frac{1}{n}$) only guarantees a $\frac{1}{n}$ approximation.}

In Theorem~\ref{theorem: general upper bound divisible items} we prove that given an instance of divisible items with reference family $\curlG$, there is always a fractional allocation that achieves a $\sfrac{n}{\sum_{G\in \curlG}{\abs{G}}}$ approximation to the $PS$. This implies the following  existential results:
\begin{itemize}
    \item For reference family that is laminar of height $h$, there exists a $\frac{1}{h}$-approximating fractional allocation.
    \item If the reference family is 2-hierarchical, there exists a  $\frac{1}{\log n}$-approximating fractional allocation.
\end{itemize}

Theorem~\ref{theorem: divisible matching negative bound} proves matching negative bounds for these three cases even with some restrictions on the valuation functions. For general reference family a $O\prs{\frac{1}{n}}$ bound on the best approximation ratio is proved, even when valuations are \emph{scaled} or \emph{dichotomous}. For laminar and 2-hierarchical families, bounds of $O\prs{\frac{1}{h}}$ and $O\prs{\frac{1}{\log n}}$ (resp.) are proven, even when all valuations are scaled.


Searching for settings with divisible items in which strong positive results do hold, we prove in Theorem~\ref{theorem: full PS laminar and resrticted} that whenever the reference group is laminar and all valuations are \emph{restricted assignments}, there exist an allocation that gives each reference group the entire \emph{proportional share}

Our first result for indivisible items, stated in Theorem~\ref{theorem:impossiblity_indivisible}, is that with no restriction on the structure of the reference family, there is no guarantee for {\bf any positive approximation ratio}, even if all agents have {\uniform} valuation functions. That is, there exist an instance with indivisible items in which all valuation functions are \unif, such that for every integral allocation $A$, there is a reference family $G\in \curlG$ with positive \emph{anyprice share} and $w_G\prs{A}=0$.
Theorem~\ref{theorem: hirarchical_impossiblity_indivisible} establishes that even if we restrict the reference family to be 2-hierarchical, no approximation better than $O\prs{\frac{1}{\sqrt n}}$ can be guaranteed, even if all valuations are \emph{identical}.


Searching for settings with indivisible items in which strong positive results do hold, we prove in Theorem~\ref{theorem: BOBW laminar and dichotomus} that whenever the valuations are dichotomous and the reference family is laminar, there exists an allocation that gives every group $G\in \curlG$ at least its entire \emph{anyprice share} $APS\prs{G}$. Moreover, there exists a randomized poly-time algorithm that always outputs such an allocation, and in expectation achieves the entire \emph{proportional share}.

{Our results when the items are \emph{goods} are summarized in Table~\ref{table:goods}.}

In Section~\ref{subsection: truthfulness results} we consider new avenues for possible manipulations by agents, introduced by our modeling of altruism. Specifically, the type of manipulations that we consider is in the choice of group to report (rather than the choice of valuation function to report). We show in claim~\ref{claim: no_truthful_good_approx} that any algorithms that operates on divisible (resp. indivisible) item instances and ensure an approximation ratio larger than $6/n$ to the \emph{propositional share} (resp. \emph{anyprice share}) for any instance with laminar family and dichotomous valuations,  cannot be \emph{truthful with respect to group reporting}.

Results for the case where the items are \emph{chores} (have negative value, or equivalently, positive costs) are presented in Section~\ref{subsection: chores results}.
When dealing with chores, if a chore $e\in \items$ has an agent $i\in \agents$ for whom $v_i\prs{e}=0$, then $e$ can be allocated to $i$ at no cost, and hence $e$ can be discarded from the instance.
Therefore, we only consider strictly positive {costs}.
Consequently, unlike the case for goods, in any instance of indivisible chores and {protected groups}, there is an integral allocation with finite approximation ratio to the \emph{proportional share} (and therefore also to the \emph{anyprice share}). We provide the following lower bounds for the possible approximation ratio in Claims~\ref{claim: lower_bound_chores_indivisible} and~\ref{claim: lower_bound_chores_divisible}: For indivisible chores no approximation ratio (to the APS) better than $\Omega\prs{\sqrt{n}}$ can be guaranteed, and for divisible chores the best ratio (to the PS) is bounded by $\Omega\prs{n}$. Both bounds hold even if valuations are scaled and the reference family is 2-hierarchical. 
Similarly to the case of goods, we observe that when all valuations are identical, there is always a fractional allocation which ensures that no group receives more than its PS. Following this observation, we conclude that when the chores are indivisible, if all agents have {\uniform} valuations and the reference family is laminar, then there is an efficient randomized algorithm that provides a ``best of both worlds" result -- every group gets a bundle of cost no larger than its APS, and in expectation (over possible bundles allocated to it), of value not larger than its PS. 
{The case of indivisible chores and identical valuation is discussed in Theorems~\ref{theorem:ChoresIdentica} and~\ref{theorem:2approxChoresIdenticaLaminar}. For arbitrary reference family {with $\groupsNum = n$ groups}, an integral allocation that $O\prs{\frac{\ln \ln n}{\ln n}}$-approximates the APS always exists, and there is a randomized efficient algorithm that produces such an allocation with high probability. Moreover, in expectation over the possible outputs of the algorithm, each group gets no more than its PS. The $O\prs{\frac{\ln \ln n}{\ln n}}$ bound is best possible (up to constant factors), even if all valuations are {\unif}. 
If the reference family is restricted to be laminar, then Theorem~\ref{theorem:2approxChoresIdenticaLaminar} establishes that there is an integral allocation that provides 2-approximation to the APS, and this allocation can be found efficiently.

{Our results regarding \emph{chores} are summarized in Table~\ref{table:chores}.}
}

Comparing between our results for goods and for chores, we find a qualitative difference in the results when items are indivisible and valuations are identical. For chores, the approximation ratios in such cases are significantly better than those for goods.


\begin{table}[hbt!]
{
\begin{subtable}[c]{0.5\textwidth}
\centering
\begin{tabular}{|c|c|c|c|}

\hline
 \diagbox{$V$}{$\curlG$} & general & laminar &2-hierarchy \\
\hline
 \specialcell{general /\\scaled} & $\Theta\prs{\frac{1}{n}}$ & $\Theta\prs{\frac{1}{h}}$  & $\Theta\prs{\frac{1}{\log n}}$   \\
\hline 
 \specialcell{restricted\\ assignment/\\dichotomous} & $\Theta\prs{\frac{1}{n}}$ & $1$ & $1$ \\
\hline
 \specialcell{identical /\\\uniform} & $1$ &$1$  & $1$  \\
\hline
\end{tabular}

\subcaption{approx. ratio to the $PS$ for divisible goods}
\end{subtable}
\begin{subtable}[c]{0.5\textwidth}
\centering
\begin{tabular}{|c|c|c|}

\hline
 \diagbox{$V$}{$\curlG$} & general &\specialcell{laminar /\\ 2-hierarchy}\\
\hline
 \specialcell{general /\\scaled /\\restricted \\assignment/\\identical} & $0$ & $O \prs{\frac{1}{\sqrt{n}}}$ \\
\hline 
 \specialcell{dichotomous/\\\uniform} & $0$  & $1$ \\
\hline
\end{tabular}
\subcaption{approx. ratio to the $APS$ for indivisible goods}
\end{subtable}

\caption{Approximation ratios for \emph{goods}. Rows indicate specific types of valuation functions, columns indicate specific types of reference families.
}

\label{table:goods}

\bigskip
\bigskip
\bigskip
\bigskip

\begin{subtable}[c]{0.5\textwidth}
\centering
\begin{tabular}{|c|c|}

\hline
 \diagbox{$V$}{$\curlG$} & \specialcell{general / laminar/ \\2-hierarchy}\\
\hline
 \specialcell{general /\\ scaled} & $\Omega\prs{n}$ \\
\hline 
 \specialcell{identical /\\ \uniform} & $1$\\
\hline 
\end{tabular}
\subcaption{approx. ratio to the $PS$ for divisible chores}
\end{subtable}
\begin{subtable}[c]{0.5\textwidth}
\centering
\begin{tabular}{|c|c|c|}

\hline
 \diagbox{$V$}{$\curlG$} & general &\specialcell{laminar /\\ 2-hierarchy}\\
\hline
 \specialcell{general /\\ scaled} & $\Omega\prs{\sqrt{n}}$ & $\Omega\prs{\sqrt{n}}$ \\
\hline 
 identical & $\Theta\prs{\frac{\log n}{\log \log n}}$ & \specialcell{${\leq 2}$\\${> 1}$}  \\
\hline
 \specialcell{\uniform} & $\Theta\prs{\frac{\log n}{\log \log n}}$  & $1$ \\
\hline
\end{tabular}
\subcaption{approx. ratio to the $APS$ for indivisible chores, {with $n$ protected groups}}
\end{subtable}
\caption{Approximation ratios for \emph{chores}. Rows indicate specific types of valuation functions, and columns indicate specific types of reference families.
}
\label{table:chores}
}
\end{table}

\subsection{Related work}

\subsubsection*{Related Item Allocation Study}

We describe previous work on fair item allocation that involves, or may be applied to, fairness towards groups of agents. Most such work considers only disjoint groups of agents, while others (as we do) consider groups that may overlap. Some works focus on \emph{envy-free} based notion of fairness, while others (as we do) consider \emph{share} notions. Most work (and ours as well) requires items to be allocated to the individual agents, but some previous work considers ``public" items that are shared by all members in the group.


There are three main sub-fields which are relevant to our research. The first one is the research of \emph{Item Allocation to Agents with Different Entitlements}. This line of study involves item allocation instances, typically of indivisible items, in which every agent might have an arbitrary entitlement to the items. Our setting, when all the reference groups in $\curlG$ are disjoint, is a special case of item allocation to agents with arbitrary entitlements -- each group in $\curlG$ in our setting is represented by a single agent $i_G$ in the arbitrary entitlements setting.  The agent $i_G$ has entitlement of $\sfrac{\abs{G}}{n}$ and an additive valuation function which is $v_G$. In ~\cite{WMMS_def} the items are indivisible and the authors introduce the fairness notion of \emph{weighted maximin share} (WMMS) for such instances. For additive valuations, they prove a tight bound of $\frac{1}{n}$-approximation to the WMMS, and showed that a $\frac{1}{2}$ approximation exists if no agent has an item which they value more than her WMMS.
The notion of \emph{any price share} (APS, the one considered in our work) was introduce in ~\cite{BEF21}. For additive valuations, the authors show that there are instances in which the entire APS cannot be guaranteed, and prove that there is always an allocation that $\frac{3}{5}$-approximates the APS, and such an allocation can be computed efficiently. While the last two mentioned papers have considered ``share based" notions as fairness criteria (WMMS and APS), as we do in our research, other work have studied other types of fairness notions.
A weighted version of \emph{envy freeness up to one good} called \emph{weighted
envy-free up to one item}  (WEF1) is studied in~\cite{WEF1_def}, in which they prove, among other results, the existence and efficient computability of a WEF1 allocation in every instance with additive valuations. All results in these works apply to our instances whenever  $\curlG$ forms a partition of (some subset of) $\agents$. In particular, since we adapt $APS$ as our fairness criterion for indivisible items, we get that if $\curlG$ forms a partition of (some subset of) $\agents$, a $\frac{3}{5}$-approximating allocation exist and can be computed efficiently. Therefore our research focuses on cases where groups in $\curlG$ may overlap.

Another related item-allocation research area is the one of \emph{Partition-Fair Item Allocation}. Usually in this type of research, the problem at hand is an item allocation instance, accompanied with a {\bf partition} of the agents to disjoint groups. The objective, similar to our work, is 
to produce allocations that are fair towards the groups in the partition. However, the details of the definitions are usually quite different (in the allocation definition, fairness criteria, structure of the groups, group welfare and valuation). 
In some of such works (e.g.~\cite{MMS_groups_shared_items, MMS_groups_shared_items_asymptotic}) the items are indivisible and allocated to the groups (and not the individual agents), where in each group $G$ every agent receives a value from the entire bundle allocated to $G$. 
{In ~\cite{Divisible_shared_within_group} the authors consider a single continuous divisible good (``cake cutting") that is to be divided to the disjoint groups (similarly to~\cite{MMS_groups_shared_items, MMS_groups_shared_items_asymptotic}). They consider the \emph{proportional share} criteria for every group, where the group valuation is the arithmetic mean of the valuations within the group, and show that a proportional (fractional) allocation exist.}
Other works explore a setting that is closer to ours, in which items are allocated to the individual agents. In~\cite{indivisible_unit_demand_types} the items are indivisible and the agents are partitioned into disjoint group $G_1,...,G_k$. The welfare group $G$ receives from a bundle is defined equivalently to ours. The items are allocated to individual agents, but only allocations that maximize welfare within each group are considered, that is first the items are partitioned into $k$ bundles $B_1,\dots, B_k$ and then each bundle $B_j$ is partitioned to the agents in $G_j$ in a way that maximizes the welfare of the group $G_j$. The objective is to eliminate envy between groups (up to one good), were envy between two groups is defined according to their welfare. In ~\cite{individual_and_group_ef} the authors consider allocations (in the same setting of indivisible items and partition of the agents) that both eliminate envy between individual agents and between groups in the partition, where the envy between groups is measured using the same WEF definition as in~\cite{WEF1_def}, taking the group sizes as weights. Among other results, they show that if all agents have the same valuation function, then there always exists an allocation that is both \emph{envy free up to any good} (EFX) for the individual agents and WEF1 between groups. If in every group all agents have the same valuation function (but valuations can differ between groups) then {there is an allocation which is both} EF1 for the individuals and WEF1 for the groups.

Finally, the much stricter notion of general \emph{Group-Fairness} requires that an allocation produced on an item-allocation instance (without any groups) will be fair for all possible subsets (sometimes only of some fixed sizes) of agents simultaneously.
Usually these works define a group-fair allocation as one that is fair (under some envy-free fairness definition) for every pair of subgroups $S,T$. 
For a single divisible item,~\cite{Devisible_Equal_Size_Group_Fairness} introduced a group-fairness notion, defined using all pairs of subgroups of equal size. 
Considering indivisible items and any pair of subgroups,~\cite{All_Groups_Fairness} define a \emph{group-fair} (GF) allocation $A$  in the following way: Denoting $A_S = \cup_{i\in S}A_i$, then $A$ is GF if for all $S,T$ there is no other allocation $A'$ such that $A'_S=A_T$ and the vector $\prs{\sfrac{\abs{S}}{\abs{T}}\cdot w_i\prs{A'}}_{i\in S}$ \emph{Pareto Dominates} the vector $\prs{w_i\prs{A}}_{i\in S}$. That is, there is no other allocation $A'$ such that $A'_S=A_T$ and for every $i\in S: \sfrac{\abs{S}}{\abs{T}}\cdot w_i\prs{A'} \geq w_i\prs{A}$ with at least one inequality being strict. The authors proceed to define two relaxations of the form ``up to one good per agent" named GF1-After (GF1A) and GF1-Before (GF1B) and proved that there is always an allocation satisfying both GF1A and GF1B, and can be computed in pseudo-polynomial time using computation of some locality based variant of a \emph{Nash Optimal} allocation. Aziz and Rey~\cite{All_Groups_Fairness_goods_and_chores} have extended GF to the case when some of the items might be chores, and presented a slightly different relaxation in the spirit of ``up to one good".
Alexander and Walsh~\cite{All_Groups_EF} introduced a weaker notion of group fairness for every two possible group sizes $k, h$, denoted as \emph{$\prs{k,h}$-Group Envy-Freeness} ($GEF_{k,h}$). They say that an allocation $A$ is $GEF_{k,h}$, if for every pair of groups $S,T$ of sizes $k,h$ (resp.) the group $S$ does not envy $T$, where $S$ utility from $A$ is defined as the arithmetic mean of utility across all agents in $S$, and $S$ utility from the bundle allocated to $T$ is defined as $\frac{1}{k\cdot h}$ times the sum of all utilities each agent in $S$ receives from the {\bf entire} bundle of $T$. They present a full taxonomy of implications between any pairs of $k,h$ and in particular, they prove that $GEF_{1,1}$ (which is simply EF) implies $GEF_{k,h}$ for any $k,h$. As immediate positive consequences, many existence and compatibility results in these works apply to out setting, moreover, they can be achieved in a \emph{group-reporting truthfulness} manner (as we can discard the original group reporting). For example, for every instance $\mathcal{I} = \prs{\agents,\items,v_1,\dots,v_n,\curlG}$ there is an allocation that is both GF1A and GF1B towards the groups in $\curlG$, and there is a truthful (w.r.t. group reporting) algorithm that produces such an allocation in pseudo-polynomial time. 
The GF1A and GF1B notions do not imply any substantial \emph{anyprice share} approximation in our setting, as they are attainable in instances in which the anyprice share approximation is not (see for example Theorem~\ref{theorem:impossiblity_indivisible}).

{
\subsubsection*{Algorithmic Fairness}

The field of \emph{algorithmic fairness} wishes to combat unfairness and biases against protected populations that result from the use of algorithms, and in particular, algorithms guided by artificial intelligence and machine learning paradigms.

For the task of \emph{prediction} in the area of \emph{machine learning}, fairness towards a specific given set of (possibly overlapping) subgroups of participating individuals is considered. These works mainly focus on combating discrimination against protected populations (e.g. minorities) in traditional and novel Machine-Learning algorithms. 
For more information see e.g.~\cite{ML_overlapping_groups, ML_multicalibration} and references therein.

Kearns et al.~\cite{gerrymandering} consider the risk of {what they call \emph{fairness gerrymandering}}, 
in many of such \emph{group-fair} prediction algorithms. Suppose we desire a prediction algorithm that will not discriminate according to some sensitive attributes (such as race, gender, age). One can define groups according to these attributes, and require that the outcome of the prediction algorithm will be fair towards each of these groups. It might happen that an output will be fair towards all predefined groups, but discriminate against some structured subgroups, defined using the very same sensitive attributes. For example, a prediction algorithm can be fair towards the group of all black individuals, as well as towards all participating females, but still might discriminate against all black woman. Such considerations do not arise in our setting.
We only require the allocation to be fair towards all groups in $\curlG$, and do not require it to fair towards groups of the form $G_1 \cap G_2$, unless the group $G_1 \cap G_2$ is explicitly present in $\curlG$.

Some \emph{allocation} tasks (that do not involve overlapping groups) have been considered as well in this literature, usually seeking some \emph{ex-ante} type of fairness. See for example~\cite{piif, mathcingFairness, LearningAllocationFairnessGroups}.

}

\subsubsection*{Altruism Definition}

Most research on item allocation assumes that every agent is selfish, in the sense that she wishes to benefit herself alone. While this assumption makes sense in many cases, especially when the agents are not human (e.g. computing tasks competing over CPU time), in other scenarios in real life, as well as some Behavioural Economics research, this assumption appears to be violated. 
Plenty research and case studies on games like the Ultimatum game (e.g.~\cite{ultimatum_game_exp_1, ultimatum_game_exp_2}), the Dictator game (e.g.~\cite{dictator_exp_1, dictator_exp_2}) and variants on those (e.g.~\cite{trust_game_Heritability}) have shown that many people tend to show altruism toward others.
Day-to-day examples exhibiting altruism include people offering financial support to family members or friends in need, or donations to charity.
Possibly, some or even all of these altruistic {behaviors} can be attributed to selfish motivations (we might tip our waiter because we want good service the next time we come back to the restaurant). Nevertheless, the \emph{observed} situation has an altruistic behaviour that needs to be taken into account.

One of our motivations for the notion of protected groups is as a way of enabling altruism in allocation problems, where an altruistic agent may declare some reference group (other than herself) as protected, and then the allocation mechanism needs to be ``fair" towards the protected group.  Other types of altruistic behavior and definitions have been studied in other Game-Theoretic settings. In Routing Games for instance, Chen and Kempe~\cite{routing_altruism} have defined an altruism level $\alpha_i \in [-1,1]$ for every player $i$. The altruistic cost of player $i$ for some outcome is composed of $\alpha_i$ times the overall social cost, plus his own egoistic social cost for that outcome. While their definition allows for different levels of altruism (i.e. values of $\alpha_i$) and even spitefulness (i.e. negative values of $\alpha_i$), it does not allow for an agent to address her altruism towards any group other than the entire set of agents. Our definition allows flexibility in choosing the reference groups, but does not allow any flexibility in the altruism level. A combinations of both can be the subject for interesting future research.

A form of altruism closer to ours appears in~\cite{social_context_games}, in which the authors study the existence of Nash equilibria in \emph{resource selections} games that are embedded in a social context in the following way: In addition to the game (defined by the set of resources $M$ and the cost function for each player $c_i$) we have also an undirected social network graph $G=\prs{N,E}$ defined on the agents. Then, each player $i$ has an aggregated cost function $f_i$ that given a cost profile $a_1,\dots,a_n$ and the social network $G$, the aggregated cost of $i$ is $f_i\prs{G,a_1,\dots,a_n}$. The authors focus on several aggregation functions, all of which depend only on the $Neigh\prs{i} = \curprs{i}\cup\curprs{j\in N: \curprs{i,j}\in E}$. In particular, among other aggregation functions, they consider the maximum, minimum and average costs within the neighborhood $Neigh\prs{i}$. The social network in their setting can be viewed as inducing reference groups{, formed using the internal group voting procedure,} with the property that every agent $i$ is included in her own reference group, and every agent $i$ must include in her group all agents that have included $i$ in their group (as $G$ is undirected). The aggregation functions also bare some similarities to ours, although in our setting we have two such functions in every instance - a valuation function $v_G$ (which we have defined as the maximum of valuations in $G$) and the welfare function $w_G$ from an allocation (which we have defined as the social welfare in $G$ ).
For other definitions of altruism in Game Theory see a recent survey by Rothe~\cite{altruism_survey} and references therein.

\subsection{Notation and terminology}

Unless obvious from context, we shall use the following notation and terminology:
\begin{itemize}
    \item We call the individual participants \emph{agents} except when they might be strategic, in which case we shall use the term \emph{players}. $\agents$ denotes the set of agents and $n$ denotes their number. We denote specific agents by the letters $i,j$ (e.g. $i \in \agents$). 
    \item $\items$ denotes the entire bundle of items. Individual items are denoted by the letters $e, f$, e.g. $e \in \items$
    \item Reference families are denoted by scripted capital letters, e.g. $\groups$, $\curlH$. {The number of groups is denoted by $\groupsNum$.}
    \item {Protected (reference)} groups are denoted by capital letters $G, H$, e.g. $G\in \groups$.
    \item Integral allocations are denoted by capital letters $A,B,C$. Bundles in a particular (integral) allocation are denoted by the allocation name, indexed by the respective agent name. For example, $A_3$ is the bundle in $A$ allocated to agent number 3. $A_3$ is a subset of $\items$.
    \item Fractional allocations are denoted by capital letters $X,Y,Z$. Fractional bundles in a particular (fractional) allocation are denoted by the allocation name, indexed by the respective agent name. For example, $X_3$ is the fractional bundle assigned to agent 3 in the fractional allocation $X$, where each entry $X_{3,e}$ is the fraction agent 3 got from item $e$. When there is only a single item, we omit 
    the item index. For example $Y_4$ will the portion allocated to agent 4, from the sole item.
    
\end{itemize}

\section{Detailed results and their proofs}

\subsection{Preliminaries}


{
In some places throughout this work, we use probabilistic arguments (randomized algorithms, the Probabilistic Method). Within those arguments, we make use of some previously known probabilistic bounds.


{
\begin{theorem*}[Chernoffs Bound]
Let $X_1,X_2,\dots, X_\ell$ be $\ell$ independent random variables that take values in $\brs{0,1}$). Let $X = \sum_{i}X_i$. For any $\mu \geq \mathbb{E}\brs{X}$ and $\delta > 0$ it holds that,
\begin{equation*}
    \Pr\brs{X > \prs{1+\delta}\mu} < \prs{\frac{e^\delta}{\prs{1+\delta}^{1+\delta}}}^{\mu}
\end{equation*}
\end{theorem*}

In addition, we make use of the following facts regarding the APS, all of which are proven in~\cite{BEF21} (we replace the budget $b$ with a group $G$ of budget $\frac{\abs{G}}{n}$).  


\begin{fact}\label{fact:apsVSps}
The \emph{anyprice share} is {a more relaxed notion} than the \emph{proportional share}, that is:
\begin{itemize}
    \item{\bf For \emph{goods}:} For any group $G\subseteq \agents$ it holds that $APS\prs{G}\leq PS\prs{G}$.
    \item{\bf For \emph{chores}:} For any group $G\subseteq \agents$ it holds that $APS\prs{G}\geq PS\prs{G}$.
\end{itemize}
\end{fact}

\begin{corollary}\label{coro:apsVSpsIntegers}
If valuations of all agents within a group $G$ are integer valued ($\forall i\in G, e\in \items: v_i\prs{e}\in \mathbb{N}$), then {for goods $APS\prs{G}\leq \floor{PS\prs{G}}$ and for chores $APS\prs{G}\geq \ceil{PS\prs{G}}$}. 
\end{corollary}

\begin{fact}\label{fact:apsMaxChores}
For chores, the APS of a group is at least the largest value the group assigns to a chore, that is $APS\prs{G}\geq \max_{e\in \items}v_G\prs{e}$
\end{fact}
}
}

\subsection{Divisible Items}\label{subsection: divisible items}
{
{
\begin{proposition}\label{prop: Divisible LP}
Given an instance $\prs{\agents,\items,V,\curlG}$ of divisible items, one can find a fractional allocation $X$ that best approximates the \emph{proportional share} in polynomial time ({in $n$, $m$ and $|\groups|$}).
\end{proposition}
\begin{proof}
Given an instance $\prs{\agents,\items,V,\curlG}$, we can find the fractional allocation guaranteeing the largest fraction of the proportional share using the following (fractional) linear program, defined on $n\cdot m + 1$ variables:
\begin{align*}
        \text{max} \quad \rho& \\
        \text{s.t.} \\
        &\forall e\in \items: \sum_{i=1}^n X_{i,e} \leq 1 \\
        &\forall G\in \curlG: \sum_{i\in G}\sum_{e\in \items} v_i(e) \cdot X_{i,e} \geq \rho \cdot PS\prs{G} \\
        &\forall i\in \agents, e\in \items: X_{i,e} \geq 0
\end{align*}

As 
linear programs can be solved in polynomial time, 
such a fractional allocation can be found in polynomial time ({in $n$, $m$ and $|\groups|$}).
\end{proof}
}


However, the question that we are concerned with is not that of efficiently finding per instance the best allocation, but rather what fraction of the proportional share does such an optimal allocation guarantee to every group. 
}
\begin{theorem}\label{theorem: general upper bound divisible items}
There is a fractional allocation that $r$-approximates the PS for all groups, where $r \doteq \frac{n}{\sum_{i=1}^{\groupsNum}{\abs{G_i}}}$.
\end{theorem}

\begin{proof}
From additivity of the $PS$ (w.r.t. disjoint bundles), it suffices
to prove the claim when there is
only a single item. 
Let $v_j$ denote the value of the item to agent $j$, and let $v_{G_p}$ denote the value of the item to group $G_p$ (namely, to the agent in $G_p$ that values the item most).
We additionally use the following notation.

\begin{align*}
    g_p &:= \abs{G_p}\\
    y_{p} &:= r\cdot \frac{g_p}{n}\\
    j_p &:= \arg\max_{j\in G_p} v_j \explanation{\text{chosen arbitrarily if there are several}}\\
\end{align*}
Now, define the fractional allocation (of the single item) to be 
$X_j := \sum_{p:j_p = j}{y_p}$. Observe that each group $G_p$ contributes exactly a 
$y_p$ fraction to the entire allocation, as $G_p$ contributes only to one agent $j_i$. 
Consequently, 
\begin{align*}
    \sum_{j=1}^{n} X_j &=\\
                        &=\sum_{p=1}^{\groupsNum}y_p\\
                        &=\sum_{p=1}^{\groupsNum}r\cdot \frac{g_p}{n}\\
                        &=\frac{r}{n}\cdot \sum_{p=1}^{\groupsNum} \abs{G_p}\\
                        &= 1 \explanation{\text{definition of }r}
\end{align*}
Hence, this allocation is legal.
The approximation factor comes from direct computations:
\begin{align*}
    \forall k\in \agents: w_{G_k}\prs{X_1,\dots X_n} &=\\
    &=\sum_{j\in G_k} X_j \cdot v_j\\
    &=\sum_{j\in G_k} \sum_{p:j_{p} = j}{y_p} \cdot v_j\\
    &=\sum_{j\in G_k} \sum_{p:j_{p} = j}{r\cdot \frac{g_{p}}{n}} \cdot v_j\\
    &\geq \sum_{p:j_p = j_k}{r\cdot \frac{g_{p}}{n}}\cdot v_{j_k} \explanation{j_k \in G_k}\\
    &\geq {r\cdot \frac{g_{k}}{n}}\cdot v_{j_k} \explanation{ \text{obviously for }k: j_k=j_k }\\
    &= r\cdot \frac{\abs{G_k}}{n}\cdot v_{G_k} \explanation{v_{j_k} = v_{G_k}}\\
    &= r \cdot PS\prs{G_k}
\end{align*}
\end{proof}

\begin{remark}\label{remark: precise upper general, laminar, hirarchical}
For ease of notation, in the last proof we assume that no two groups in the family are identical. This assumption can be made without loss of generality, because is some of the groups are identical, we only need to accommodate one of them. That is, suppose that after deleting multiplicities the groups are $G_{i_1},\dots ,G_{i_t} $. Then there is always an allocation with approximation factor of $r \doteq \frac{n}{\sum_{s=1}^{t}{\abs{G_{i_s}}}}$.\\
{
Theorem~\ref{theorem: general upper bound divisible items} implies that:
\begin{itemize}
    \item For a laminar reference family of height $h$ there is always an allocation with approximation factor $r_2 = \sfrac{1}{h}$.
    \item For a 2-hierarchical reference family there is always an allocation with approximation factor $r_3 = \sfrac{1}{\log n}$.
\end{itemize}
}
\end{remark}

{The following theorem shows that the upper bounds in {Remark \ref{remark: precise upper general, laminar, hirarchical} are best possible} (up to a constant factor).}

\begin{theorem}\label{theorem: divisible matching negative bound}
The upper bounds in {Remark \ref{remark: precise upper general, laminar, hirarchical} are best possible} (up to a constant factor) in the following cases: 
\begin{itemize}
{
    \item For general reference family - no approximation ratio better than $O\prs{\frac{1}{n}}$ can be guaranteed, even if all valuations are {scaled} or dichotomous.
    \item For laminar reference family of height $h$ - no approximation ratio better than $O\prs{\frac{1}{h}}$ can be guaranteed, even if all valuations are {scaled}.
    \item For 2-hierarchical family - no approximation ratio better than $O\prs{\frac{1}{\log n}}$ can be guaranteed, even if all valuations are {scaled}.
    }
\end{itemize}

\end{theorem}

\begin{proof}

{\bf General reference family and {scaled} valuations}.

{
Suppose there are only two items, and the valuations are
\begin{align*}
    v_i\prs{1} &= 
        \begin{cases}
            1-\frac{1}{n^2} & \text{if $i\leq \sfrac{n}{2}$} \\
            1-\frac{1}{n}  & \text{else} 
        \end{cases} \\
    v_i\prs{2} &= 
        \begin{cases}
            \frac{1}{n^2} & \text{if $i\leq \sfrac{n}{2}$} \\
            \frac{1}{n}   & \text{else} 
        \end{cases} \\
\end{align*}
Notice that the valuations are indeed {scaled} (for very agent, the value for the entire bundle is exactly 1).
Let the reference groups be $\curprs{G_i = \left\{i, \frac{n}{2}+1,...,n\right\}: i=1,\dots \frac{n}{2}}$. Clearly \[\forall i\in \brs{\frac{n}{2}}:  PS\left(G_i\right) = \left(\frac{1}{2} + \frac{1}{n}\right)\cdot\prs{1-\frac{1}{n^2} + \frac{1}{n}}\] 

Consider an arbitrary fractional allocation $X$. Denote the portion of item $1$ that is given to agent $i$ by $X_{i,1}$. As there are $n$ agents and $\sum_i X_{i,1} \leq 1$,  there must be $i\in \brs{\frac{n}{2}}$ with $X_{i,1} \leq \frac{2}{n}$. For such $i$, it must be that $w_{G_i}\prs{X} \leq \frac{2}{n}\cdot \prs{1-\frac{1}{n^2}}+ \frac{1}{n}$ (as at best the second item is distributed only among agents $\frac{n}{2}+1,\dots,n$ which would yield a total profit of $\frac{1}{n}$ for $G_i$). Let us bound the ratio $\frac{w_{G_i}\prs{X}}{PS\left(G_i\right)}$ and get a bound for the approximation ratio.
\begin{align*}
    \frac{w_{G_i}\prs{X}}{PS\left(G_i\right)} &=\\
                                            &=\frac{w_{G_i}\prs{X}}{\prs{\frac{1}{2} + \frac{1}{n}}\cdot\prs{1-\frac{1}{n^2} + \frac{1}{n}}} \\
                                            &\leq \frac{\frac{2}{n}\cdot \prs{1-\frac{1}{n^2}}+ \frac{1}{n}}{\prs{\frac{1}{2} + \frac{1}{n}}\cdot\prs{1-\frac{1}{n^2} + \frac{1}{n}}}\\
                                            &= O\prs{\frac{1}{n}}
\end{align*}

}

{\bf  General reference family and dichotomous valuations.}

{
Suppose there are $n$ agents and $m=n$ items $e_1,\dots, e_n$(where $n$ is even). Let the valuations be
\begin{align*}
                 &v_i\prs{e_1} = 1 \\
    \forall j>1: &v_i\prs{e_j} = 
        \begin{cases}
            1 & \text{if $i\leq \sfrac{n}{2}$} \\
            0  & \text{else} 
        \end{cases} \\
\end{align*}
Notice that this valuation are indeed \emph{dichotomous}. Suppose that the 
reference groups are\\ $\curprs{G_i = \left\{i, \frac{n}{2}+1,...,n\right\}: i=1,\dots \frac{n}{2}}$. Clearly \[\forall i\in \brs{\frac{n}{2}}:  PS\left(G_i\right) = \left(\frac{1}{2} + \frac{1}{n}\right)\cdot n\] 

Consider an arbitrary fractional allocation $X$. Denote the portion of item $e_j$ that is given to agent $i$ by $X_{i,j}$. 
As $X$ is legal, than $\forall e{_j}\in M: \sum_i X_{i,j} \leq 1$. Summing over $j=2,...,m$ we get:
\begin{align*}
    \sum_{j=2}^{n} \sum_{i\in [n]} X_{i,j} \leq n-1
\end{align*}
Therefore there is an $i\in \brs{\frac{n}{2}}$ with $\sum_{j=2}^{n}X_{i,j} \leq \frac{2}{n}\cdot \prs{n-1}=2-\frac{2}{n}$. For this $i$, it must be that $w_{G_i}\prs{X} \leq \prs{2-\frac{2}{n}}\cdot 1 + 1$ (as at best the first item is distributed only among agents $i, \frac{n}{2}+1,\dots,n$ which would yield a total profit of $1$ for $G_i$, and agent $i$ got at most total fraction of $2-\frac{2}{n}$ from all the items).
Let us bound the ratio $\frac{w_{G_i}\prs{X}}{PS\left(G_i\right)}$ and get a bound for the approximation ratio.
\begin{align*}
    \frac{w_{G_i}\prs{X}}{PS\left(G_i\right)} &=\\
                                            &=\frac{w_{G_i}\prs{X}}{\left(\frac{1}{2} + \frac{1}{n}\right)\cdot n} \\
                                            &\leq \frac{\prs{2-\frac{2}{n}}\cdot 1 + 1}{\left(\frac{1}{2} + \frac{1}{n}\right)\cdot n}\\
                                            &= O\prs{\frac{1}{n}}
\end{align*}

}

{\bf Laminar reference family and {scaled} valuations.}

{
For laminar families, the approximation ratio in theorem \ref{theorem: general upper bound divisible items} and its following remark, takes the form of $\sfrac{1}{h}$, where $h$ denotes the family height (the longest chain of strict containment). In order to prove a matching lower bound, we introduce for any large enough $h$, an instance with reference family of height $h$, and prove that no allocation yields a better approximation than $O\prs{\sfrac{1}{h}}$.

Let $h\geq 2$ be an arbitrary positive integer. We consider an example with
$m = 2^h$ items, and $n \simeq h2^h$ agents. 
The agents are partitioned into $h + 1$ classes, $C_0,\dots,C_h$. For $0 \leq t < h$, class $C_t$ has $2^{h-t}$ agents, and each agent
has value $2^{-t}$ for $2^t$ distinct items (no two agents of the same class have an
item that they both value). For example, $C_0$ has $2^h$ agents, and each agent
has value 1 for a different item. Class $C_1$ has $2^{h-1} = \frac m 2$ agents, and each agent has
value $\frac 1 2$ for two distinct items. Class $C_h$ is different. It contains $h\cdot 2^h$ agents
(rather than just one agent), and each agent has value $2^{-h}$ for each of the
items.
Therefore the total number of agents is $n = h2^h + \sum_{t=0}^{h-1}2^{h-t} = h2^h + 2\prs{2^h-1}$. As we only consider $h\geq$2, we get that $n\leq h2^h + h\prs{2^h-1}\leq 2\cdot h\cdot 2^h$. This means that the class $C_h$ alone, which has $h2^h$ agents, contains at least half of the agents. 
Observe that all valuation functions are {scaled} (the sum of values is 1).

Let the reference groups (after deleting multiplicities) be $G_j = \bigcup_{t=j}^{h} C_t$ for every $0 \leq j \leq h$.
Observe that this family is indeed a laminar of height $h$ (as $\forall j=0,\dots,h-1: G_j \subsetneq G_{j+1}$).
Each group $G_j$ contains the class $C_h$ which in turn holds at least half of all the agents.
Hence the $\forall j=0,\dots,h: PS\prs{G_j} \geq \frac{1}{2}\cdot 2^h \cdot 2^{-j} = 2^{h-j-1}$, by giving all items to agents in class $C_j$ .
Consider an arbitrary fractional allocation $X$ and suppose that $\forall j: w_{G_j}\prs{X}\geq \rho \cdot PS\prs{G_j}$. From the lower bound on $PS\prs{G_j}$  we get that $w_{G_j}\prs{X} \geq \rho \cdot 2^{h-j-1} $. 
For $0 \leq t \leq h$, let $y_t$ denote
the (perhaps fractional) number of items allocated to agents in class $C_t$ . Then
$\sum_{t} y_t \leq m=2^h$. The welfare of the group $G_j$ can be written as $w_{G_j}\prs{X}=\sum_{t=j}^{h}y_t 2^{-t}=\sum_{t=0}^{h-j}y_{j+t} 2^{-j-t}$.
Then for every group $j$ we have 
\begin{align*}
    &\sum_{t=0}^{h-j}y_{j+t} 2^{-j-t} \geq \rho \cdot 2^{h-j-1}\\
    \iff&\sum_{t=0}^{h-j}y_{j+t} 2^{-t} \geq \rho \cdot 2^{h-1}\\
    \iff&2^{1-h}\sum_{t=0}^{h-j}y_{j+t} 2^{-t} \geq \rho\\
    \iff&\frac{2}{m}\cdot \sum_{t=0}^{h-j}y_{j+t} 2^{-t} \geq \rho\\
\end{align*}

Summing over all groups we have that
\begin{align*}
    (h + 1)\rho &\leq \\
    &\leq \sum_{j=0}^{h} \frac{2}{m}\cdot \sum_{t=0}^{h-j}y_{j+t} 2^{-t} \\
    &=\frac{2}{m} \sum_{j=0}^{h} \sum_{t=0}^{h-j}y_{j+t} 2^{-t}\\
    &\leq \frac{2}{m} \cdot 2 \cdot \sum_{j=0}^{h}y_t \\
    &\leq \frac{2}{m} \cdot 2 \cdot m\\
    &= 4\\
    \implies& \rho \leq \frac{4}{h+1}\\
\end{align*}
}

{\bf 2-hierarchical family and {scaled} valuations.}

{
Let $n$ be large enough integer, such that both $\frac{n}{\log n } \text{ and } \frac{\log n}{2}$ are integers (in particular $\log n \geq 2$) and $\frac{1}{\sqrt{n}} \geq \frac{\log n }{n}$. Let $m=\frac{n}{\log n }$ be the number of items, and denote $D=\frac{\log n}{2}$.
The $n$ agents are partitioned into $D+1$ classes $C_0,C1,\dots,C_D$. Classes $C_0,\dots, C_{D-1}$ have $m$ agents each. The last class $C_D$ contains the rest $\frac{n}{2}$ agents.
We refine this partition by dividing each class $C_d$ into $2^d$ sub-classes $C^{1}_{d}\dots C^{2^d}_{d}$, each one has $\frac{\abs{C_d}}{2^d}$ agents.\\
For $d<D$, in each subclass $C^{s}_d$, each agent has value of $2^{-d}$ for $2^d$ distinct items (this subclass is of size $\frac{m}{2^d}$). No two agents of the same
subclass $C^{s}_d$ have an item that they both value.
The classes $C^{s}_D$ are different - each agent $i\in C_D = \bigcup C^{s}_D$ has value $\frac{1}{m}$ for each item. Observe that all valuations are indeed {scaled}.
As consequence of this partition, we get that 
\begin{align*}
    \forall e\in \items: v_{C^{s}_{d}}\prs{e} = 
    \begin{cases}
        2^{-d} & d<D \\
        \frac{1}{m} & d=D \\
    \end{cases}
\end{align*}
Next, we define the reference family, using a binary tree $T$ of depth $D$ defined on the sub-classes in the following way:
The nodes are all the sub-classes in the partition, where in level $d$ the nodes are all the sub-classes $C^{1}_d,\dots,C^{2^d}_d$ (the root is $C^1_0$ and the leaves are $C^{1}_D\dots C^{2^D}_{D}$). For each internal (non-leaf) node $C^s_d$, its two children nodes are $C^{2s-1}_{d+1}, C^{2s}_{d+1}$. 
Using this tree, we define the reference groups. For every internal node of the tree $C^s_d$, let $G^s_d$ be the group containing all sub-classes that are in the sub-tree rooted at $C^s_d$ (including $C^s_d$ itself). The number of groups is $2^D - 1  < \sqrt{n}$ so this can be realized in our model {even with internal group voting}. This family is also 2-hierarchical, as $\curprs{C^s_d: d=0,\dots,D; s=1,\dots,2^d}$ is a partition of the agents, and for all $d$ all the sub-classes $C^s_d$ are of the same size.
Observe that for each group $G^s_d$, it's valuations function is identical to $v_{C^{s}_{d}}$, as the root sub-class $C^{s}_{d} $ of the associated sub-tree has the highest value for every item (here we are using the assumption that $2^{-D} = \frac{1}{\sqrt{n}} \geq \frac{\log n }{n} = \frac{1}{m}$).
We claim that no fractional allocation $X$ can give approximation ratio to $PS$ better than $\frac{8}{\log n}$. First observe that it suffices to prove this claim for allocations that distribute all of the items to the first $D$ classes, as every group containing a sub-class $C^s_D$, contains also the sub-class $c^{\floor{\sfrac{s}{2}}}_{D-1}$ that values every item at least as much as the sub-class $C^s_D$.
Denote $G_d = \bigcup_{s=1}^{2^d}G^s_d$ ($G_d$ is not part of the reference family). Then, from additivity of $PS$ w.r.t disjoint groups that have the same valuation function, we get that every allocation $X$ that achieves an approximation ratio of $\rho$, also guarantees that $\forall d\in \curprs{0,\dots,D-1}: w_{G_d}\prs{X}\geq \rho \cdot PS\prs{G_d}$. Let us compute what is $PS\prs{G_d}$. First, the size of this group is 
\begin{align*}
    \abs{G_d} &=\\
              &= \sum_{j=d}^{D}\abs{C_j}\\
              &= n - \sum_{j=0}^{d-1}\abs{C_j}\\
              &= n - \sum_{j=0}^{d-1}m\\
              &= n - dm\\
              &= n\prs{1-\frac{d}{\log n}}
\end{align*}
The value this group has for any item, is identical to the value assigned to it by the top-most class in $G_d$, i.e. $v_{G_d}\prs{\items} = v_{C_d}\prs{\items} = m\cdot 2^{-d}$. Therefore {$PS\prs{G_d} = m\cdot 2^{-d}\prs{1-\frac{d}{\log n}}$}.
With a slight abuse of notation, denote $X_d$ to be the total amount of items given to the class $C_d$ ($X_d$ can be non-integer as the allocation can be fractional). The welfare of the group $G_d$ can be then written as $w_{G_d}\prs{X} = \sum_{j=d}^{D-1} X_j\cdot 2^{-j}$. With this notation, the optimal approximation ratio $\rho$ (and an associated allocation $X$), can be found using the following linear program $\mathcal{P}$, defined on the $D+1$ variables $X_0,\dots,X_{D-1},\rho$:
\begin{align*}
        \text{max} \quad \rho& \\
        \text{s.t.} \\
        &\sum_{d=0}^{D-1} X_d = m \\
        & \sum_{j=d}^{D-1} 2^{-j}\cdot X_j \geq \rho \cdot PS\prs{G_d} & \forall d\in \curprs{0,\dots,D-1} \\
        & X_d \geq 0&\forall d\in \curprs{0,\dots,D-1}
\end{align*}

Or equivalently,
\begin{align*}
        \text{max} \quad \rho& \\
        \text{s.t.} \\
        &\sum_{d=0}^{D-1} X_d = m \\
        & \rho \cdot PS\prs{G_d} + \sum_{j=d}^{D-1}-2^{-j}\cdot X_j \leq 0  & \forall d\in \curprs{0,\dots,D-1} \\
        & X_d \geq 0&\forall d\in \curprs{0,\dots,D-1}
\end{align*}
The dual program $\mathcal{D}$ is the following program, defined of the $D+1$ variables $Y_0,\dots,Y_{D-1},Z$:
\begin{align*}
        \text{min} \quad m\cdot Z& \\
        \text{s.t.} \\
        &\sum_{d=0}^{D-1} PS\prs{G_d} \cdot Y_d = 1 \\
        & Z + \sum_{j=0}^{d} -2^{-d} \cdot Y_j  \geq 0 &\forall d\in \curprs{0,\dots,D-1} \\
        & Y_d \geq 0&\forall d\in \curprs{0,\dots,D-1}
\end{align*}
Or equivalently,
\begin{align*}
        \text{min} \quad m\cdot Z& \\
        \text{s.t.} \\
        &\sum_{d=0}^{D-1} PS\prs{G_d} \cdot Y_d = 1 \\
        &2^{-d} \cdot \sum_{j=0}^{d} Y_j  \leq Z &\forall d\in \curprs{0,\dots,D-1} \\
        &Y_d \geq 0&\forall d\in \curprs{0,\dots,D-1}
\end{align*}
From the \emph{weak primal-dual} Theorem, to show an upper bound $B$ for $\rho$, it suffices to show a feasible solution $y_0,\dots,y_{D-1},z$ to $\mathcal{D}$ such that $m\cdot z = B$.
Let $z=\frac{8}{n}$ and $y_d=\frac{1}{D\cdot PS\prs{G_d}}$.
Obviously $\forall d\in \curprs{0,\dots,D-1}, y_d \geq 0 $ and by definition $\sum_{d=0}^{D-1} PS\prs{G_d} \cdot y_d = \sum_{d=0}^{D-1} PS\prs{G_d} \cdot \frac{1}{D \cdot PS\prs{G_d}} = 1$. It remains to show that $\forall d\in \curprs{0,\dots,D-1}: 2^{-d} \cdot \sum_{j=0}^{d} y_j  \leq z$
or equivalently $\forall d\in \curprs{0,\dots,D-1}: \sum_{j=0}^{d} \frac{1}{PS\prs{G_d}} \leq \frac{4\log n \cdot2^{d}}{n}$.
Indeed:
\begin{align*}
    \sum_{j=0}^{d} \frac{1}{PS\prs{G_d}} =&\\
    &=\sum_{j=0}^{d} \frac{1}{m\cdot 2^{-j} \cdot \prs{1-\frac{j}{\log n}}}\\
    &=\frac{1}{m}\sum_{j=0}^{d} \frac{2^j}{\prs{1-\frac{j}{\log n}}}\\
    &=\frac{\log n}{m}\sum_{j=0}^{d} \frac{2^j}{\log n -j}\\
    &=\frac{\log^2 n}{n}\sum_{j=0}^{d} \frac{2^j}{\log n -j}\\
    &\leq \frac{\log^2 n}{n}\sum_{j=0}^{d} \frac{2^j}{\log n - \frac{\log n}{2}} &\explanation{j\leq D < \frac{\log n}{2} }\\
    &= \frac{2\log n}{n}\cdot \sum_{j=0}^{d} 2^j \\
    &=  \frac{2\log n}{n}\cdot\prs{2^{d+1}-1}\\
    &\leq \frac{2\log n}{n}\cdot\prs{2^{d+1}}\\
    &=\frac{4\log n}{n}\cdot 2^d
\end{align*}

Therefore this is a feasible solution to the program $\mathcal{D}$. Its value is $m\cdot z = \frac{n}{\log n} \cdot \frac{8}{n} = \frac{8}{\log n}$, which shows that the primal program $\mathcal{P}$ has no feasible solution with value larger than  
$\frac{8}{\log n}$
}
\end{proof}

\begin{theorem}\label{theorem: full PS laminar and resrticted}
If all valuations are \emph{restricted assignments} and the reference groups form a \emph{laminar family} then it is always possible to give each group at least its full proportional share.
\end{theorem}

\begin{proof}
As in Theorem \ref{theorem: general upper bound divisible items}, from additivity of the proportional share (w.r.t. disjoint bundles), it suffices to prove the claim when there is only a single item.
Let $\curlG$ be the reference family, and assume that it is laminar. We extend $\curlG$ to a possibly larger laminar family $\curlH$. First, in addition to every group $G\in \curlG$, we add the entire set of agents $\agents$ to $\curlH$. Next, we add to $\curlH$ every ``missing" group from $\curlG$: For a group $G\in \curlG$ denote the (possibly empty) set of its direct ``children" groups in hierarchy by  $C\left(G\right)$. That is, $C\prs{G} := \curprs{G'\in \curlG: G'\subsetneq G \text{ and } \forall G''\in \curlG: G'\subsetneq G'' \rightarrow G \subseteq G''}$. As $\curlG$ is laminar, $C\left(G\right)$ is a collection of disjoint subsets of $G$. We complete it to a full cover of $G$ by adding to $\curlH$ the group $S\left(G\right) := G \setminus \bigcup _{G' \in C\left(G\right)} G'$. Formally, the new family is:
\[
\curlH := \curlG \cup \curprs{N} \cup \left\{S\left(G\right): G\in \curlG \right\} 
\]
It is easy to see that $\curlH$ remains laminar.
We claim that there is always a fractional allocation in which every group in $\curlH$ - and therefore every group in $G \in \curlG$ - gets at least its entire proportional share.
As there is only a single item, we denote the valuations of this item simply by $v_i$. Valuations are in form of \emph{restricted share}, meaning there is a number $w > 0$ s.t. $\forall i \in \agents, v_i \in \curprs{0, w}$. If $\forall i \in \agents: v_i = 0$ then we are done (no group has a positive proportional share, so any allocation will do). From now on we assume that this is not the case. Throughout the remainder of the proof, we shall use the following definitions and notations:
\begin{itemize}
    \item Say that an agent $i$ \emph{wants} the item if $v_i = w$.
    \item Say that a group $G$ \emph{wants} the item if there is an agent $i\in G$ that wants the item (and therefore $v_G = w$).
    \item For an agent $i\in \agents$ we define its \emph{heir} group  $H(i) \in \curlH$ to be the minimal (w.r.t. inclusion) group s.t. $i\in H(i)$ and $H(i)$ wants the item.
    \item For an agent $i\in \agents$ we define its \emph{heir} agent  $h(i) = 
    \begin{cases}
       i & \text{if $i$ want the item}\\
       \min \curprs{k\in H(i): k \text{ wants the item}} & \text{else}
    \end{cases}
    $
\end{itemize}
Note that of $H(i)$ always exist since $\agents\in \curlH$ and we assume that some agent wants the item. This implies that $h(i)$ is well defined. Also note that if $k=h(i)$ for some $i$, then $h(k)=k$ as $k$ wants the item. Moreover, if $i\in H \in \curlH$ and $H$ wants the item, then $h(i) \in H$. Indeed if $i$ wants the item it is obvious, and if not then $h(i)$ is in the minimal group within $\curlH$ which contains $i$ and want the item. As $H$ is such a group and $\curlH$ is laminar, $h(i) \in H' \in \curlH$ and $H' \subseteq H$.
\par Our allocation $X$ is the following - we start with an even distribution allocation $X'$ (that is $\forall i: X'_i = \sfrac{1}{n}$) and each agent passes along it's share to it's heir. Formally 
\[
X_i := \frac{\left|h^{-1}(i)\right|}{n}
\]
Notice that since $h(i)$ is well defined, $h^{-1}$ forms a partition of the set of agents $\agents$. Hence $X$ is a legal solution, since,
\begin{align*}
  \sum_{i\in \agents} X_i &= \sum_{i\in N}  \frac{\left|h^{-1}(i)\right|}{n} \\
                    &= \frac{1}{n} \cdot \sum_{i\in \agents}\left|h^{-1}(i)\right| \\
                    &= \frac{1}{n} \cdot n \explanation{\text{$h^{-1}$ partitions $\agents$}} \\
                    &= 1
\end{align*}

We claim that this allocation provides each group in $H \in \curlH$ at least its entire proportional share.
Fix $H\in \curlH$. If $H$ does not want the item, we are done (as $PS\left(H\right) = 0$ so any allocation will do). Else we know that $\forall k \in H: h(k) \in H$, so 
\begin{align*}
             & \bigcup_{i \in H} {h^{-1}(i)} \supseteq H\\
    \implies & \sum _{i \in H} {\left|h^{-1}(i)\right|} \geq \left|H\right| \explanation{h^{-1}(i_1) \cap h^{-1}(i_2) = \emptyset}\\
    \implies & \sum _{i \in H} {X_i} \geq \frac{\left|H\right|}{n}\\
    \implies & \sum _{i \in H} {X_i \cdot v_H} \geq w\cdot \frac{\left|H\right|}{n} \explanation{H \text{ want the item}}\\
    \implies & w_H\left(X\right) \geq PS\left(H\right)\\
\end{align*}
\end{proof}


\subsection{Indivisible Items}\label{subsection: indivisible items}



\begin{theorem}\label{theorem:impossiblity_indivisible}
With general reference family, no positive approximation to the  \emph{anyprice share} can be guaranteed, even with {\uniform} valuations.
\end{theorem}
\begin{proof}
{
Our proof uses the following \emph{Hitting-Set} argument.

\begin{lemma}\label{lemma: hitting-set argument}
For any large enough and even $n$, there exist a family $\curlG \subset 2^{\brs{n}}$ s.t.
\begin{itemize}
    \item $\abs{\curlG} \leq n$
    \item $\forall G\in \curlG: \abs{G}={\frac{n}{2}} $
    \item Every hitting set $S$ of $\curlG$ must be of size at least 3
\end{itemize}
\end{lemma}

\begin{proof}[Proof of Lemma \ref{lemma: hitting-set argument}]
We prove the lemma using the probabilistic method.
Suppose $\curlG$ is made of {$n$} sets, each chosen uniformly and independently at random from all $\binom{n}{n/2}$ subsets of size $n/2$. Denote them as $G_1,\dots {G_{n}}$. Observe that the first two conditions in the Lemma holds trivially from construction. It remains to show that with positive probability, this family has no hitting set of size less that three.


For each $i\in \brs{n}, k\in \brs{{n}}: \Pr\brs{i\in G_k} = \sfrac{1}{2}$. {For any $i\neq j$ and each group $G_k$, it holds that $\Pr\brs{\curprs{i,j}\subseteq [n] \setminus G_k} = \frac{\binom{n-2}{\sfrac{n}{2}-2}}{\binom{n}{\sfrac{n}{2}}} = \frac{n-2}{4\cdot\prs{n-1}}\geq \frac{1}{8}$}. Consequently, $\forall i,j\in [n], k\in \brs{{n}}: \Pr\brs{\curprs{ i,j}\cap  G_k \neq \emptyset} \leq {\sfrac{7}{8}}$. As the groups were chosen independently, $ \Pr\brs{\curprs{i,j}\text{ is a hitting set for } \curlG } \leq \prs{{\sfrac{7}{8}}}^{{n}}$ 
and from the union bound:
\begin{align*}
    \Pr \brs{\text{There's no hitting set of size 2 for } \curlG} &=&\\
    &=\Pr \brs{\forall \curprs{i,j}\subseteq \brs{n}: \curprs{i,j}\text{ is not a hitting set for } \curlG}&\\
    &=1-\Pr\brs{\exists \curprs{i,j}\subseteq \brs{n}: \curprs{i,j}\text{ is a hitting set for } \curlG}&\\
    &\geq 1 - \binom{n}{2}\cdot \left({\sfrac{7}{8}}\right)^{n} & \explanation{\text{union bound}}\\
    &> 0 &\explanation{\text{for any large enough } n}
\end{align*}
\end{proof}

Next, we prove the theorem itself.
Suppose $m=2$ and $n$ is even and large enough as in Lemma \ref{lemma: hitting-set argument}. According to this Lemma, there exists a reference family $\curlG$ consisting of some {$n$ groups $G_1,...,G_n$} each of size $n/2$, with no hitting set of size 2. {Suppose that all agents have the \uniform valuation function.}
As $\forall i: \abs{G_i} = \frac{n}{2}$, each group $G_i$ admits a strictly positive \emph{anyprice share}. Since there are only two indivisible items, for an allocation $A$ to be positive approximating the $APS$ there must be at most two agents $i,j$ that get the items, and every group $G_k$ must contain at least one of them (otherwise this group welfare will be 0). I.e. $\curprs{i,j}$ is a hitting set for $\curlG$, in contradiction to its choice.
{As $\abs{\groups}=n$, this example can be realized even with internal protected group voting.}
}
\end{proof}


\begin{theorem}\label{theorem: hirarchical_impossiblity_indivisible}
When the reference family is laminar or even 2-hierarchical, no approximation ratio better than $O\prs{\frac{1}{\sqrt{n}}}$ can be guaranteed, even with identical valuations.
\end{theorem}
\begin{proof}
{
Let $n\geq 4$, be an integer of the form $n=k^2$ for some $k\in \NN$.
Suppose there are $n$ items and $k = \sqrt{n}$ disjoint groups, each of size $k = \sqrt{n}$.
{All $n = k^2$ agents have the same valuation function $v$, defined as:
\begin{align*}
    v\prs{e_t} = 
    \begin{cases}
        1 & t=1,2,\dots,k-1\\
        \frac{1}{n-k+1} & t=k,k+1,\dots,n
    \end{cases}
\end{align*}
}

Hence the MMS of every agent is {$\frac{1}{n-k+1} > 0$, whereas the
MMS (and APS) of every group is $1$} (the MMS partition for a group has $k-1$ bundles
with a single large item, and one bundle with all small items). We remark
that in this specific instance, the MMS, APS and TPS are all equal to each
other.
{Next, we describe the formation of the groups using an internal voting process.} 
Suppose that the reference groups are voted to be the following:
The first $k-1$ agents are egoistic (report themselves), and agent $k$ reports the group $\curprs{1,2,\dots,k}$.
Next, agents $k+1,k+2,...,2k-1$ are also egoistic, and agent $2k$ reports the group $\curprs{k+1,k+2,\dots,2k}$, and so on.
The reference groups formed are:
\begin{align*}
    G_i = \begin{cases}
        \curprs{i-k+1,i-k+2,\dots,i} & i\mod k \equiv 0 \\
        \curprs{i} & \text{else}
    \end{cases}
\end{align*}
Notice that this structure forms a $\sfrac{1}{\sqrt{n}}$-hierarchy, which is also a 2-hierarchy for $n\geq 4$.

In every integral allocation that positively approximates the APS, every egoistic agent must
receive at least one item, as its APS is strictly positive. In one of the larger groups, w.l.o.g $G_k$, agents receive only
small items (as there are more large groups than large items). We claim that the total number of (small) items received by members of $G_k$ is at most $2k-1$. To that end, recall that there are only $n$ items, from which $n-2k+1$ are given to the $n-2k+1$ egoistic agents that are not in $G_k$. This leaves only $2k-1$ items to group $G_k$. Hence the group $G_k$ receives a value of at most {$\prs{2k-1}\cdot \frac{1}{n-k+1} < \frac{2}{k-1}$, whereas $APS\prs{G_k} = 1$. This means that the approximation ratio is bounded by $\frac{2}{k-1}=\frac{2}{\sqrt{n}-1}$}.
}
\end{proof}

\begin{theorem}\label{theorem: BOBW laminar and dichotomus}
If all valuations are dichotomous and the reference family is laminar, then it is always possible to give each group at least its entire \emph{anyprice share}. Moreover, there is a poly-time randomized algorithm, that always yields such an allocations, and promises in expectation each group its entire \emph{proportional share} (``best of both worlds" algorithm).
\end{theorem}

Following Theorem \ref{theorem: full PS laminar and resrticted}, there exists a fractional allocation $X$ which grants each group it's entire \emph{proportional share}. We exploit a previous result by {Budish, Che, Kojima and Milgrom} (\cite{10.1257/aer.103.2.585}) {which implies that $X$} can be rounded to an integral allocation that satisfies our requirements.
\begin{claim}\label{claim: budish et. al.}\cite{10.1257/aer.103.2.585}
Let $\curlH\subseteq2^{\agents\times \items}$ be a family of sets that contains all singletons ${\curprs{\prs{i,e}}, i\in \agents, e\in \items}$. For a legal allocation (fractional or integral) $X$ and a set $S\in 2^{\agents\times \items}$, denote $X_S = \sum_{\prs{i,e}\in S} X_{i,e}$ (for integral allocation $X_{i,e} \in \curprs{0,1}$). If $\curlH$ can be partitioned into two laminar families $\curlH_1, \curlH_2$, then for any legal fractional allocation $X$, there exist positive numbers $\alpha_1\dots \alpha_K$ with $\sum_{k\in \brs{K}} \alpha_k = 1$ and allocations $A^1,\dots,A^K$ s.t.
\begin{enumerate}
    \item $\sum_{k=1}^{K} \alpha_k \cdot A^k = X$
    \item $\forall S\in \curlH, k\in [K]: A^k_{S}\in \curprs{\floor{X_{S}},\ceil{X_{S}}}$
\end{enumerate}
Moreover, there is a poly-time randomized algorithm, that always yields one of the allocations $\curprs{A^1\dots A^K}$, and produce allocation $A_k$ with probability $\alpha_k$.
\end{claim}

In the next corollary we show how to fit any fractional allocation to this framework, provided that the reference family is laminar.
\begin{corollary}\label{coro: rounding fractional to integral with laminar}
Let $I = \prs{\agents,\items,V,\curlG}$ be an allocation instance {with protected groups}. For a legal allocation (fractional or integral) $X$ and a set $S\in 2^{\agents}$, denote $X_S = \sum_{i \in S} \sum _{e \in \items} X_{i,e}$. If $\curlG$ is laminar, then for any legal fractional allocation $X$, there is a $K\in \mathbb{N}$, positive numbers $\alpha_1\dots \alpha_K$ with $\sum_{k\in \brs{K}} \alpha_k = 1$, and integral legal allocations $A^1,\dots,A^K$ s.t.
\begin{enumerate}
    \item $\forall i\in \agents,e\in \items: \sum_{k=1}^{K} \alpha_k \cdot A^k_{i,e} = X_{i,e}$ \label{cond1}
    \item $\forall G\in \curlG, k\in [K]: A^k_{G}\in \curprs{\floor{X_{G}},\ceil{X_{G}}}$\label{cond2}
    \item $\forall i\in \agents, k\in [K]: (A^k)_{i,e} = 1 \implies X_{i,e} > 0$\label{cond3}
\end{enumerate}
Moreover, there is a poly-time randomized algorithm, that always yields one of the allocations $\curprs{A^1\dots A^K}$, and produce allocation $A^k$ with probability $\alpha_k$ 
\end{corollary}

\begin{proof}
We simply fit this setting, into the setting of Claim \ref{claim: budish et. al.}. Define the following:
\begin{align*}
    \curlH_1 &:= \curprs{\curprs{\prs{i,e}}: i\in \agents, e\in \items}\cup \curprs{\agents\times \curprs{e}: e\in \items}\\
    \curlH_2 &:= \curprs{G_i \times \items: G_i\in \curlG}\\
    \curlH &:= \curlH_1 \cup \curlH_2
\end{align*}
It is easy to see that $\curlH_1$ is laminar, and since $\curlG$ is laminar so is $\curlH_2$. Claim \ref{claim: budish et. al.} promises a poly-time algorithm that produces an allocation $A^k$ w.p. $\alpha_k$, and that \begin{enumerate}
    \item $\sum_{k=1}^{K} \alpha_k \cdot A^k = X$ \label{eq1: sum_k a_k A^k = X}
    \item $\forall S\in \curlH, k\in [K]: A^k_S\in \curprs{\floor{X_{S}},\ceil{X_{S}}}$ \label{eq2: A^k_S in floor ceil}
\end{enumerate}
Now, since for any $i\in \agents, e\in \items$ the singleton $\curprs{\prs{i,e}}$ is in $\curlH$, then \ref{eq2: A^k_S in floor ceil} implies that $A^k_{i,e} \in \curprs{\floor{X_{i,e}},\ceil{X_{i,e}}}$, but $X$ is legal, so $X_{i,e} \in \brs{0,1}$ meaning $A^k_{i,e} \in \curprs{0,1}$.
Similarly, since for any item $e\in \items$, the set $\agents\times \curprs{e}$ is in $\curlH$, we get that $\sum_{i\in \agents} A^k_{i,e} = A^k_{\agents\times \curprs{e}} \in \curprs{\floor{X_{\agents\times \curprs{e}}},\ceil{X_{\agents\times \curprs{e}}}}$. Again, since $X$ is legal, we know that $X_{\agents\times \curprs{e}} = \sum_{i\in \agents} X_{i,e} \leq 1$, so we get that also ${\sum_{i\in \agents} A^k_{i,e} \leq 1}$. Both results ensures that $A^k$ is indeed a legal integral allocation.

\par Obviously equation \ref{eq1: sum_k a_k A^k = X} implies that $\forall i\in \agents, e\in \items \sum_{k=1}^{K} \alpha_k \cdot A^k_{i,e} = X_{i,e}$. The fact that for any $G \in \curlG$, the set $G \times \items$ is in $\curlH$, combined with equation \ref{eq2: A^k_S in floor ceil}, ensures that $A^k_{G}\in \curprs{\floor{X_{G}},\ceil{X_{G}}}$. Finally, equation \ref{eq2: A^k_S in floor ceil} along with the fact that $\curlH$ contains all singletons $\curprs{\prs{i,e}}$ guarantees that if item $e$ is given to agent $i$ in some allocation $A^k$, then $X_{i,e} > 0$ as well.
\end{proof}
Using this corollary, we next prove Theorem \ref{theorem: BOBW laminar and dichotomus}

{
\begin{proof}[Theorem \ref{theorem: BOBW laminar and dichotomus} proof]
Consider the fractional allocation $X$ {that best approximates the $PS$. Proposition~\ref{prop: Divisible LP} 
implies that 
$X$ can be found in polynomial time. As assured by Theorem~\ref{theorem: full PS laminar and resrticted}, this allocation {guarantees} the entire proportional share to any group $G \in \curlG$.} We simply run the algorithm from Corollary \ref{coro: rounding fractional to integral with laminar} on $X$, and denote its output allocation $A$. The polynomial run-time of this algorithm is given by Corollary \ref{coro: rounding fractional to integral with laminar}.

\par {As valuations are dichotomous, 
Corollary~\ref{coro:apsVSpsIntegers} implies for every $G \in \curlG$  that  $APS\prs{G} \leq \floor{PS\prs{G}}$.} Following promise no. \ref{cond2} we get that $A_{G}\geq APS\prs{G}$. We are not done, since this only shows that the number of items that $G$ got is at least $APS\prs{G}$, but not which agents got these items. However, promise no. \ref{cond3} implies that any item $e$ that is allocated to $i'\in G$ satisfies $X_{i',e} > 0$. From the {optimality of $X$}, this implies that $v_{i'}\prs{e} = 1$. So the group $G$ got {at least} $\floor{PS\prs{G}}$ items, and each such item contributes $1$ to $w_{G}\prs{A}$, concluding $w_{G} \geq \floor{PS\prs{G}} \geq APS \prs{G}$.
\par Finally, for any group $G$ it holds that

\begin{align*}
    \mathbb{E}\brs{w_{G}\prs{A}} &=\\
    &= \sum_{k=1}^{K} \alpha_k \cdot w_{G}\prs{A^k}\\
    &= \sum_{k=1}^{K} \alpha_k \cdot A^k_{G}  \explanation{\forall i \in G, \forall e\in A_{i}: v_{i}\prs{e} = 1 \text { from eq. \ref{cond3}}}\\
    &= \sum_{k=1}^{K} \alpha_k \cdot  \sum_{i\in G}\sum_{e\in \items}A^k_{i,e}\\
    &= \sum_{i\in G}\sum_{e\in \items}\sum_{k=1}^{K} \alpha_k \cdot A^k_{i,e}\\
    &=\sum_{i\in G}\sum_{e\in \items}X_{i,e} \explanation{\text{equation } \ref{cond1}}\\
    &\geq PS\prs{G} \explanation{\text{since $X$ promises the $PS$, and all valuations are dichotomous}}
\end{align*}
\end{proof}
}

\subsection{Truthful Group Reporting}\label{subsection: truthfulness results}


{
\begin{claim} \label{claim: no_truthful_good_approx}
Let $\mathcal{X}$ be an allocation algorithm that operates on instances with divisible items, dichotomous valuations and laminar families. If $\mathcal{X}$ guarantees a $\rho>\frac{6}{n}$ approximation to the \emph{proportional share}, then it cannot be truthful.
The statement remains true after replacing \emph{divisible items} with \emph{indivisible items} and  \emph{proportional share} with \emph{anyprice share}.

\end{claim}

\begin{proof}
First, we shall prove the claim regarding divisible items. Assume towards contradiction the existence of a truthful algorithm $\mathcal{X}$ that for every instance of divisible items, dichotomous valuation and laminar reference family, produces an allocation of an approximation ratio at least $\rho > \frac{6}{n}$.
Consider an example with $n$ items, and $n$ players, where $n>2$ and even. The valuations are
\begin{align*}
    v_i = 
    \begin{cases}
        \Vec{1} & i\leq \frac{n}{2}\\
        \prs{1,0,0,\dots,0} & \text{else} 
    \end {cases}
\end{align*}

The true reference groups are $G_1 = \curprs{1}, G_2 = \curprs{2},\dots, G_{\sfrac{n}{2}} = \curprs{\sfrac{n}{2}}$, hence $PS\prs{G_i} = 1, \forall i=1,\dots,\sfrac{n}{2}$.
Let $X$ be the allocation produced by $\mathcal{X}$ on this instance. There must be at least one player $i\in \brs{\frac{n}{2}}$, such that $\sum_{e\in M}X_{i,e} \leq 2$. Assume w.l.o.g that this is player number 1, that is $w_{G_1}\prs{X}\leq 2$.
Now consider the same example but the first reference group is $G'_1 =  \curprs{1,\sfrac{n}{2}+1\dots,n}$ instead of $G_1$. Observe that family is still laminar, hence $\mathcal{X}$ should produce an allocation $X'$ that s.t. $w_{G'_1}\prs{X'}\geq \rho PS\prs{G'_1}$. Since all other players in $G'_1$ other that $i=1$, have zero value for any item but the first, it implies that $w_{G'_1}\prs{X'} = w_{G_1}\prs{X'} + \sum_{i>\sfrac{n}{2}}X'_{i,1} \leq  w_{G_1}\prs{X'} + 1$. As $\abs{G'_1}>\frac{n}{2}$, we get that $PS\prs{G'_1}>\frac{n}{2}$. Combining both, we obtain that $w_{G_1}\prs{X'} > \rho \cdot \frac{n}{2} - 1 > 2 \geq w_{G_1}\prs{X}$. \emph{This means} that player 1 would strictly gain by misreporting $G'_1$ instead of $G_1$, contradicting the truthfulness of $\mathcal{X}$.

To prove the claim for the indivisible case, we take the same instances as in the divisible case (except that the items are indivisible, of course). 
Assume that algorithm $\mathcal{A}$ always produces integral allocations with approximation ratio at least $\rho$. Let $A$ be the allocation that $\mathcal{A}$ produces given the true reference groups. As in the divisible case, we can assume that $w_{G_1}\prs{A} \leq 2$. Since $\abs{G'_1}\geq \frac{n}{2}$, it must be that $APS\prs{G'_1}\geq \frac{m}{2} = \frac{n}{2}$, as for any possible pricing, there are at least $\frac{m}{2}$ items with total price at most $\frac{1}{2}$, which is within the budget of $G'_1$. Denoting by $A'$ the allocation produced by $\mathcal{A}$ ($A'$ promises $\rho$ approximation as the family is still laminar), it is easy to see that $w_{G'_1}\prs{A'} \leq w_{G_1}\prs{A'} + 1$, and from the approximation ratio we derive that $w_{G_1}\prs{A'} \geq \rho \cdot APS\prs{G'_1} - 1 > \frac{6}{n}\cdot\frac{n}{2} - 1 = 2 \geq w_{G_1}\prs{A}$. 
\end{proof}
}

\subsection{Chores}\label{subsection: chores results}


The value of chore $e$ to group $G$ is defined to be $v_G\prs{e} = \min_{i\in G} v_i\prs{e}$.

\begin{claim}
{In every instance of allocation of indivisible chores to agents with additive valuation functions {and with protected groups}, there is an integral allocation in which for every group, the ratio between the value that it receives and its proportional share is finite. Moreover, such an allocation can be computed efficiently.}
\end{claim}
\begin{proof}
As we do not allow agents to value chores as $\infty$, an allocation $B$ will fail to achieve a {finite} approximation factor, only if for some group $G$, it holds that $PS\prs{G} = 0$, and this group received a positive social-cost. But the case of $PS\prs{G} = 0$ can only happen if for every chore $e$, there exist an agent $i_e\in G$, such that $v_{i_e}\prs{e} = 0$. We can define a new allocation $A$ that gives every chore $e$ to the respective $i_e$. For this allocation, $\forall H \subseteq \agents, c_{H}\prs{A} = 0 \leq PS\prs{H}$.  
\end{proof}

\begin{claim}\label{claim: lower_bound_chores_indivisible}
When chores are indivisible, there are allocation instances in which one cannot give each group less than $\Omega\prs{\sqrt{n}}$ times the $APS$, even with a $\sqrt{n}$-hierarchical family and {scaled} valuations
\end{claim}

\begin{proof}
Suppose there are $m=\sqrt{n}$ items and let $\varepsilon=\frac{1}{n^{1.5}}$.
For $i=1,\dots,\sqrt{n}$, agent $i$ has value of $\varepsilon$ for the i'th chore, and $\frac{1-\varepsilon}{m-1}$ for the rest of the chores.
All other agents have value of $\frac{1}{m}$ for any chore.
It is easy to see that all valuations are indeed {scaled}.
Let the reference groups be $G_1 = \curprs{1,\dots,\sqrt{n}}, G_2=\agents$.
The valuation function for both groups is $v_{G_1}\prs{e} = v_{G_2}\prs{e} =\varepsilon$, as $\varepsilon = \frac{1}{n^{1.5}} < \frac{1}{m} = \frac{1}{\sqrt{n}}$. The $APS$ of the two groups is substantially different. As $G_2=\agents$, this group must get all the chores, so $APS\prs{G_2}=v_{G_2}\prs{\items}=\varepsilon \cdot m$. For $G_1$, $APS\prs{G_1} = \varepsilon$, as for any adversarially set prices, there will always be at least one chore with price at least $\frac{1}{\sqrt{n}}$, and it suffices that $G_1$ takes this single chore.
Let $A$ be an integral allocation. We shall prove that it does not achieves an approximation $\alpha < \sqrt{n}$ to the $APS$. If for some agent $i>\sqrt{n}, A_i \neq \emptyset$, than $c_{G_2}\prs{A}\geq \frac{1}{m}=\frac{1}{\sqrt{n}}=\sqrt{n}\cdot\frac{1}{n^{1.5}}\cdot m > \alpha \cdot APS\prs{G_2}$, meaning the $\alpha$ factor was not achieved.
Otherwise, all chores are distributed across the first $\sqrt{n}$ agents, in which case $c_{G_1}\prs{A} \geq m \cdot \varepsilon = \sqrt{n}\cdot APS\prs{G_1}>\alpha \cdot APS\prs{G_1}$

\end{proof}

\begin{claim}\label{claim: lower_bound_chores_divisible}
When chores are divisible, one cannot give each group less than $\Omega\prs{n}$ times the $PS$, even with a $\frac{n}{2}$-hierarchical family and {scaled} valuations.
\end{claim}

\begin{proof}
Suppose that there are two items, and $n\geq 4$ agents. Let $\varepsilon = \frac{1}{2n}$. The valuation functions are,
\begin{align*}
    v_i = 
    \begin{cases}
        \prs{\varepsilon,1-\varepsilon} & i=1\\
        \prs{1-\varepsilon, \varepsilon} &i=2\\
        \prs{\sfrac{1}{2},\sfrac{1}{2}} &i>2\\
    \end{cases}
\end{align*}
We only have two groups $G_1 = \agents, G_2 = \curprs{1,2}$.
Observe that all valuations are indeed {scaled}, and the family is 2-hierarchical (even $\sfrac{n}{2}$-hierarchical). 
Let $X$ be any fractional allocation, and assume it achieves an $\alpha$ approximation. That is $\forall G \in \curprs{G_1,G_2}, c_G\prs{X}\leq \alpha PS\prs{G}$.  The $PS$ of the groups are, \begin{align*}
    PS(G_k) = 
    \begin{cases}
        2\varepsilon & k=1\\
        \frac{4\varepsilon}{n} & k=2
    \end{cases}
\end{align*}
Let $x,y$ be the fractions of chores 1 and 2 (respectively) allocated to agents $[n]\setminus \curprs{1,2}$. The cost of $G_1$, is then at least $c_{G_1}\prs{X} \geq \frac{x+y}{2}$, and from the approximation, we know that $c_{G_1}\prs{X}\leq 2 \alpha \varepsilon$, implying that $x+y \leq 4 \alpha \varepsilon$. 
The cost of $G_2$ is at least $c_{G_2}\prs{X} \geq \varepsilon \cdot\prs{2-\prs{x+y}}$. Again from the approximation, we know that $c_{G_2}\prs{X}\leq  \frac{4 \alpha \varepsilon}{n}$. Combining them, we get:
\begin{align*}
    &\frac{4 \alpha \varepsilon}{n} \geq \varepsilon \cdot\prs{2-\prs{x+y}}\\
    \implies &\frac{4 \alpha}{n} \geq 2-\prs{4 \alpha \varepsilon}\\
    \implies &2 \alpha \geq n-\prs{2 \alpha \varepsilon n}\\
    \implies &\alpha \geq \frac{n}{2+2n\varepsilon}\\
    \implies &\alpha \geq \frac{n}{3}\\
\end{align*}
\end{proof}

\begin{observation}\label{observ: chores divisile identical}
One can always give no more than $PS$ for divisible chores and identical valuations.
Indeed, a uniform allocation ($\forall i\in \agents, e\in \items: X_{i,e} = \frac{1}{n}$) promises any group $G\subseteq \agents$, exactly $PS\prs{G}$.
\end{observation}

\begin{corollary}
For indivisible chores with a laminar reference family and {\uniform} valuations (i.e. $\forall i\in \agents, e\in \items: v_i\prs{e} = 1$), there always is an allocation that gives every group no more than its $APS$. Moreover, there is a poly-time randomised algorithm that always returns such an allocation, and in expectation gives every group no more than its $PS$. 
\end{corollary}

\begin{proof}
Following Observation \ref{observ: chores divisile identical}, we can find a fractional allocation $X$ that guarantees that each group received no more than its $PS$. As the reference family is laminar, Corollary \ref{coro: rounding fractional to integral with laminar} provide us with a BoBW algorithm that always yield an integral allocation $A$ in which $c_{G}\prs{A} \leq \ceil{X_G} = \ceil{\frac{\abs{G} m}{n}} \leq APS\prs{G}$, and in expectation (on the algorithm randomness) the social cost of any group is $X_G = PS\prs{G}$
\end{proof}


{Next, we discuss the case when all valuations are \emph{identical}, {the number of protected groups is $n$}, and the family structure is not restricted. The following Theorem establishes that there is always an integral allocation that achieves a $O\prs{\frac{\ln n}{\ln \ln n}}$ approximation to the APS, and this bound is best possible (up to a constant factor).}

\begin{theorem}\label{theorem:ChoresIdentica}
Let $\prs{\agents,\items,\mathcal{G}, v}$ be an instance of indivisible chores allocation, with identical valuations which are all $v$, {and with $n$ protected groups}. 
\begin{itemize}
    \item There always exists an integral allocation that $O\prs{\frac{\ln n}{\ln \ln n}}$ approximates the APS. Moreover, there exists a randomized algorithm that outputs such an allocation with probability at least $1-\frac{1}{n}$, and in expectation gives each group its exact PS. 
    This algorithm disregards the reference family, and therefore it is  \emph{truthful w.r.t. group reporting}.
    \item There exist an instance with {\uniform}valuations, in which no integral allocation guarantee less than $\frac{\ln n}{\ln \ln n}$ of the \emph{anyprice share} to each group.
\end{itemize}
\end{theorem}

\begin{proof}
We start by proving the first item. 
Without loss of generality, we assume that the ${\max_{e\in M}v\prs{e} = 1}$ (otherwise we divide each value by $\max_{e\in M}v\prs{e}$). Our allocation $A$ is chosen uniformly and independently at random, meaning $\forall i\in \agents, e \in \items: \Pr\brs{e\in A_i}=\frac{1}{n}$. This allocation clearly achieves the exact $PS$ in expectation, and ignores the $\curlG$ component of the input, so it is truthful w.r.t. group reporting.
It remains to show that there exists a constant $t$ such that ${\Pr\brs{\exists G\in \curlG, c_G\prs{A}>t\cdot\frac{\ln n}{\ln \ln n}\cdot APS\prs{G}}\leq \frac{1}{n}}$. For small values of $n$ such that $\frac{\ln \ln \ln n}{\ln \ln n} >\frac{1}{2}$, any allocation will provide a constant approximation, so we may assume that $\frac{\ln \ln \ln n}{\ln \ln n} \leq \frac{1}{2}$. Let us introduce some random variables induced by $A$.
\begin{itemize}
    \item $C_{i,e} = v\prs{e}\cdot \mathds{1}_{e\in A_i}$ where $\mathds{1}_{D}$ is the indicator R.V. for the event $D$.
    \item $C_G = \sum_{i \in G}\sum_{e\in \items}C_{i,e}$
\end{itemize}
Fix a non empty group $G\subseteq \agents$. Clearly $\mathbb{E}\brs{C_G}=PS\prs{G}$. 
Denote $\mu=\max\curprs{PS\prs{G},1},{t=2\cdot e}$ and let ${\delta = t\cdot \frac{\ln n}{\ln\ \ln n}-1}$. As $C_G$ is the sum of independent random variables, all taking values in the range $[0,1]$, and as $\mu \geq \mathbb{E}\brs{C_G}$, we may use the Chernoff bound,

\begin{align*}
            \Pr{\brs{C_G > \prs{1+\delta}\mu}} 
            &\leq \prs{\frac{e^\delta}{\prs{1+\delta}^{\prs{1+\delta}}}}^{\mu}\\
            &\leq  \prs{\frac{e}{1+\delta}}^{\prs{1+\delta}\mu}\\
            &=  \prs{\frac{1}{\frac{2\ln n}{\ln \ln n}}}^{\prs{\frac{2e\cdot \ln n}{\ln \ln n}}\mu}\\
            &\leq \prs{\frac{\ln n}{\ln \ln n}^{\frac{\ln n}{\ln \ln n}}}^{-2e} \explanation{\mu \geq 1}\\
            &= \prs{n^{1-\frac{\ln \ln \ln n}{\ln \ln n}}}^{-2e}\\
            &\leq \prs{n^{\frac{1}{2}}}^{-2e} \explanation{\text{for any large enough } n}\\
            &= \frac{1}{n^e}
\end{align*}
{From Facts~\ref{fact:apsVSps} and ~\ref{fact:apsMaxChores}}, for chores it always holds that $APS\prs{G}\geq \max\curprs{{\max_{e\in \items}v_G\prs{e}, PS\prs{G}}}$. We have that $\Pr\brs{c_G\prs{A}>t\cdot\frac{\ln n}{\ln \ln n}\cdot APS\prs{G}} \leq \Pr{\brs{C_G > \prs{1+\delta}\mu}} \leq \frac{1}{n^e}$. By the union bound, we get that $\Pr\brs{\exists G\in \curlG, c_G\prs{A}>t\cdot\frac{\ln n}{\ln \ln n}\cdot APS\prs{G}}\leq n\cdot \frac{1}{n^e} < \frac{1}{n}$. 

Coming to prove the impossibility part of the theorem, we use the next lemma.

\begin{lemma}
\label{lem:randomSets}
For any large enough $n\in \mathbb{N}$ such that $k=\frac{\ln n}{\ln \ln n} \in \mathbb{N}$, there exists a family $\curlG \subseteq 2^{[n]}$ consisting of $n$ sets, each of size $\frac{n}{k}$, such that for any set $K\subseteq[n]$ of size exactly $k$ there exists $G\in \curlG$ with $K\subseteq G$.
\end{lemma}

\begin{proof}[Proof of the Lemma]
We prove the lemma using the probabilistic method. Suppose $\curlG$ is made of {$n$} sets, each chosen uniformly and independently at random from all $\binom{n}{n/k}$ subsets of size $n/k$. Fix a set $K\subseteq[n]$ of size exactly $k$. For every set $G\in \curlG$ it holds that $\Pr\brs{K\subseteq G}=\frac{\binom{n-k}{\frac{n}{k}-k}}{\binom{n}{\frac{n}{k}}}$. Form binomial coefficient properties ($\binom{n}{k} = \frac{n}{k}\binom{n-1}{k-1}$), we derive that $\Pr\brs{K\subseteq G} = \frac{\sfrac{n}{k}}{n} \cdot \frac{\sfrac{n}{k} - 1}{n - 1}\cdot \frac{\sfrac{n}{k} - 2}{n - 2} \cdots\frac{\sfrac{n}{k} - k + 1}{n - k + 1}$. 
We can lower bound this probability with $\Pr\brs{K\subseteq G} \geq \prs{\frac{\frac{n}{k}-k}{n}}^k = \prs{\frac{1}{k} - \frac{k}{n}}^k$. 
As $k^2=o\prs{n}$, we get that for any sufficiently large enough $n$ it holds that $\Pr\brs{K\subseteq G} \geq \prs{\frac{1}{2k}}^k$. The family $\curlG$ contains $n$ independently chosen such sets, implying that $\Pr\brs{\forall G \in \mathcal{G}, K \not \nsubseteq  G} \leq \prs{1 - \prs{\frac{1}{2k}}^k}^n$. From the union bound, the probability that exists such set $K$ is bounded by
\begin{equation*}
    \binom{n}{k}\cdot\prs{1 - \prs{\frac{1}{2k}}^k}^n \leq n^k \cdot \prs{1 - \prs{\frac{1}{2k}}^k}^n \xrightarrow[n \to \infty]{} 0
\end{equation*}
where the last limit holds since $k=\frac{\ln n}{\ln \ln n}$. This implies that for any sufficiently large enough $n$ with positive probability, it holds that $\forall K\subseteq [n] \text{ with } \abs{K}=k, \exists G\in \curlG \text{ s.t. } K\subseteq G$. 
\end{proof}

Completing the proof of the impossibility result, choose $n, k$ and $\curlG$ as in Lemma~\ref{lem:randomSets}, and suppose that there are $m=k=\frac{\ln n}{\ln \ln n}$ chores. Let all valuations be {\unif.} 
As there are exactly $k$ identical chores, and every group $G\in \curlG$ is of size exactly $\frac{n}{k}$ it is easy to see that $\forall G\in \curlG, APS\prs{G} = 1$. 
Let $A$ be an integral allocation, and let $K'\subseteq \agents$ be the subset of agents such that $\cup_{i\in K'}A_i = \items$ and $\forall i\in K', A_i \neq \emptyset$. As $\abs{\items}=m=k$ we have that $\abs{K'}\leq k$.

By Lemma~\ref{lem:randomSets}, 
there is $G\in \curlG$ with $K'\subseteq G$. For such a group $G$ it holds that $c_G\prs{A} = v_G\prs{\items}=k$, whereas $APS\prs{G}=1$.
\end{proof}

The next Theorem establishes that a 2-approximation allocation always exists for instances with indivisible chores and identical valuations, if the reference family is laminar. Notice that since there are instances of indivisible chores and identical valuations for which the $APS$ cannot be guaranteed to all individual agents, then a constant approximation is the best we can hope for. We leave open the question of whether an approximation ratio $\rho<2$ exists for such instances.

\begin{theorem}\label{theorem:2approxChoresIdenticaLaminar}
Let $\prs{\agents,\items,\mathcal{G}, v}$ be an instance of indivisible chores allocation with identical valuations which are all $v$, and $\curlG$ is laminar. There exists an integral allocation $A$ that 2-approximates the APS. Such an allocation can be found in polynomial time. 
\end{theorem}

\begin{proof}
We begin by extending $\curlG$ into a possibly larger laminar family $\curlH$ such that $\curlG \subseteq \curlH$. First, we add every group $G\in \curlG$ to $\curlH$, and next we add the groups $\agents$ and $\curprs{i}$ for all $i\in \agents$ to $\curlH$. Throughout the proof we interchangeably refer to groups in $\curlH$ as groups or nodes in the tree representing the hierarchy of containment in $\curlH$.

Observe that $\curlH$ remains laminar. In addition, for every group $H\in \curlH$ of size at least 2, there exist some disjoint $k\geq 2$ subgroups $H_1,\dots,H_k$ such that $\cup H_i = H$ and  $\forall i\in[k], H_i\in \curlH$.
We describe a ``trickle down" inductive procedure of distributing the chores, where each internal (non-leaf) group $H\in\curlH$ receives a bundle of chores $M_H$, which is a subset of the chores allocated to its parent group, and in turn, distributes $M_H$ to its direct subgroups. At the end of the procedure, all the chores are distributed across the leafs of $\curlH$ -- the individual agents.
For two groups $G \subset H$ and a non-empty bundle of chores $L \subseteq \items$, denote $PS_{H,L}\prs{G} = \prs{\sfrac{\abs{G}}{\abs{H}}}\cdot v\prs{L}$. (The usual $PS\prs{G}$ can be written as $PS_{\agents,\items}\prs{G}$.) 
We use this notation to describe how the bundle of a parent group $H$ is distributed to it children subgroups.
The procedure is defined by the following inductive process:

\begin{enumerate}
    \item $M_\agents=\items$: Start with all chores (the bundle $\items$) at the root -- the group $\agents$ (which is in $\curlH$ by construction). 
    \item Let $H\in \curlH$ be an internal node in the hierarchy, that has received a bundle $M_H\subseteq \items$ in the process, and needs to distribute $M_H$ to its children subgroups $H_1,\dots,H_k$ ($k\geq 2$). Let $e_H\in M_H $ be such that $v\prs{e_H} = \max_{e\in M_H}v\prs{e}$. (The choice of max is rather arbitrary here, and other choices such as min, or item of largest index, will work just as well.) Set $M'_H \leftarrow \prs{M_H\setminus \curprs{e_H}}$
    \item For $i=1,\dots,k-1$:
    \begin{enumerate}
        \item $H_i$ selects and removes chores (one by one) from $M'_H$ into $M_{H_i}$, up until the first time that either $M'_H=\emptyset$ or $v\prs{M_{H_i}}\geq PS_{H, M_H\setminus \curprs{e_H}}\prs{H_i}$.\label{distribution step}
    \end{enumerate}
    \item The group $H_k$ receives all the remaining chores in $M_H$ -- which are all the remaining chores in $M'_H$ and the chore $e_H$. 
\end{enumerate}

The procedure ends with an integral allocation $A$ (to the individual agents). The procedure runs in polynomial time, because there are at most {$2n-1$} groups in $\curlH$ {(as the tree representing $\curlH$ has $n$ leaves and every internal node has at least two children)}, and for each internal group that allocates $M_H$ to its $k$ direct subgroups, the procedure computes the allocation in time polynomial in the size of $M_H$ and $k$. 

We prove the approximation ratio using the following claim.

\begin{claim}
For every group $H\in \curlH$, either $M_H=\emptyset$ or there exist a chore $e\in M_H$ such that $v\prs{M_H\setminus \curprs{e}} \leq PS\prs{H}$
\end{claim}

\begin{proof}
We prove this claim using induction on the level of $H$ in the tree $T$ that represents the hierarchy of $\curlH$. If $H$ is at the root, that is $H=\agents$, then $M_H=\items$ and the claim holds as $\forall e\in \items, v\prs{\items\setminus\curprs{e}} \leq v\prs{\items} = v\prs{M_H} = PS\prs{\agents}$.
For the inductive step, we assume that the claim holds for some group $H$ that is not a leaf in the tree, and prove the claim for its children groups $H_1,\dots,H_k$ ($k>1$, as $H$ is internal node). Let $H_i$ be a direct child of $H$. If $M_H=\emptyset$ then also $M_{H_i}=\emptyset$ and we are done. Otherwise, by the induction hypothesis, there exists a chore $e\in M_H$ such that $v\prs{M_H\setminus\curprs{e}}\leq PS\prs{H}$, and in particular $v\prs{M_H\setminus\curprs{e_H}}\leq PS\prs{H}$ for $e_H$ such that $v\prs{e_H} = \max_{e\in M_H} v\prs{e}$. We consider two cases for the value of $i$. 
\begin{description}
    \item [\boldmath${i<k}$:] If at some point during an iteration $j<i$ we have reached $M'_H = \emptyset$ then again $M_{H_i}=\emptyset$. If this is not the case, then for the very last chore $e_i$ that was added to $M_{H_i}$ it holds that $v\prs{M_{H_i}\setminus\curprs{e_i}} \leq PS_{H, M_H\setminus \curprs{e_H}}\prs{H_i}$. That is $v\prs{M_{H_i}\setminus\curprs{e_i}} \leq \frac{\abs{H_{i}}}{\abs{H}}v\prs{M_H\setminus\curprs{e_H}}$. As $v\prs{M_H\setminus\curprs{e}_H}\leq PS\prs{H}$ we conclude that $v\prs{M_{H_i}\setminus\curprs{e_i}} \leq \frac{\abs{H_{i}}}{\abs{H}}\cdot PS\prs{H} = \frac{\abs{H_i}}{\abs{H}}\cdot \frac{\abs{H}}{n}v\prs{\items} = \frac{\abs{H_i}}{n}v\prs{\items} = PS\prs{H_i}$.
    \item[\boldmath $i=k$:] If at some point during an iteration $j<k$ we have reached $M'_H = \emptyset$ then $M_{H_k}=\curprs{e_H}$ and $v\prs{M_{H_k}\setminus\curprs{e_H}}=0\leq PS\prs{H_k}$. If this is not the case, then for all $j<k$ it holds that $v\prs{M_{H_j}} \geq PS_{H, M_H\setminus \curprs{e_H}}\prs{H_j} = \frac{\abs{H_j}}{\abs{H}}v\prs{M_H\setminus \curprs{e_H}}$. Therefore, as $\abs{H} - \sum_{j<k}\abs{H_i} =\abs{H_k}$ we have that $v\prs{M_{H_k}\setminus\curprs{e_H}} \leq \frac{\abs{H_k}}{\abs{H}}v\prs{M_H\setminus \curprs{e_H}}\leq \frac{\abs{H_k}}{\abs{H}} PS\prs{H} = PS\prs{H_k}$. 
\end{description}
\end{proof}

Since $\curlH$ is laminar, and as all valuations are identical, for every group $H \in \curlH$ we have $c_H\prs{A} = v\prs{M_H}$ -- the laminarity assures that $M_H$ is not changed in the rest of the process, and from identical valuations we get that however $M_H$ is distributed across the agents in $H$, the cost of $H$ from the resulting allocation remains the same.
Following the last claim, this implies that $\forall H\in \curlH$ either $c_H\prs{A}=0\leq PS\prs{H}$ or $\exists e\in M_H: c_H\prs{A} \leq PS\prs{H} + v\prs{e}$. 
Facts~\ref{fact:apsVSps} and ~\ref{fact:apsMaxChores} imply that $APS\prs{H} \leq PS\prs{H}$ and $APS\prs{H} \leq \max_{e\in \items} v_H\prs{e}$, and as for every $H\in\curlH$ with $M_H\neq \emptyset$ for every chore $e\in  M_H: v\prs{e} \leq \max_{e'\in \items}v\prs{e'}$, we conclude that either $c_H\prs{A}=0\leq APS\prs{H}$ or $c_H\prs{A} \leq PS\prs{H} + \max_{e\in \items}v\prs{e} \leq 2\cdot APS\prs{H}$. 
Since $\curlG \subseteq \curlH$ we get that the 2-approximation holds for every group $G\in \curlG$.
\end{proof}

\section{Discussion, open questions, possible extensions}


In this work, we study the problem of fairly allocating items to agents 
{when there are protected groups}.
To that end, we have defined the valuation function of group $G$ in a way that captures the essence of {benefiting the group as a whole}. 
That is, the value of a bundle $S$ of items to $G$ is the maximum welfare that $G$ can derive from the bundle by allocating the items of $S$ in an optimal way among the members of $G$. (For goods, this takes the form of $\max$, and for chores it takes the form of $\min$ cost.) We consider both divisible and indivisible items, and for both we adapt previously known \emph{share} based notions of fairness, namely \emph{proportional share} for divisible items, and \emph{anyprice share} for indivisible items. 

{
Given some of our negative results, one might wonder whether they can be circumvented by using a different definition for the group valuation, one that 
is less demanding than our welfare maximizing definition, and hence easier to satisfy. 
A natural requirement from such a definition is that it should ``comply to consensus". 
Namely, if all agents within a group have the same valuation function $v$, then the group's valuation function is also $v$.
Our definition of group valuation functions ($\max_{i\in G}v_i\prs{e}$ for goods, $\min_{i\in G}v_i\prs{e}$ for chores) complies to consensus, but there are alternative definitions that do as well, such as the average $\frac{1}{\abs{G}}\cdot \sum_{i\in G} v_i\prs{e}$. Alas, the proofs Theorems~\ref{theorem:impossiblity_indivisible} and~\ref{theorem: hirarchical_impossiblity_indivisible} that show negative results for indivisible goods hold when all valuations are identical. Hence the \emph{anyprice share} will remain the same for every such ``complying to consensus" group valuation function, and the negative results in these theorems will hold. For divisible goods however, taking for example the average to be the group valuation function, one can see that the balanced distribution allocation ($X_{i,e} = \frac{1}{n}$) guarantees every group $G\subseteq \agents$ its exact $PS$ (see also~\cite{Divisible_shared_within_group}).

{
One may wonder how our results will differ if in addition to the {protected groups}, we are required to be fair towards all individuals as well. That is, in every instance we have $\curlG=\curlG_1\cup\curlG_2$, where $\curlG_1$ are the {protected} reference groups, and $\curlG_2 = \curprs{\curprs{i}:i\in \agents}$. Obviously all our negative results still hold. For our positive results, whenever a $\rho = \Theta\prs{1}$ approximation was obtainable, the same constant $\rho$ remains obtainable for this extension, and whenever a non-constant $\rho$ approximation was achievable, a $\Theta\prs{\rho}$ approximation is still achievable. For example, the extended reference families in Theorems~\ref{theorem: full PS laminar and resrticted} and~\ref{theorem:2approxChoresIdenticaLaminar} either already include, or can be extended to include at no cost, all individual agents. Note that whenever the $\curlG_1$ is laminar of height $h$, $\curlG$ is laminar of height at most $h+1$, and if $\curlG_1$ is 2-hierarchical than so is $\curlG$. If $\sum_{G\in\curlG_i}\abs{G} = s\cdot n$ then $\sum_{G\in\curlG}\abs{G} \leq (s+1)\cdot n$. Therefore the bounds in Remark~\ref{remark: precise upper general, laminar, hirarchical} do hold, up to constant factors.
}.


}
\subsection*{Acknowledgements}

{Research supported by the Israel Science Foundation (grant number 5219/17).}

\bibliographystyle{plain}
\bibliography{references}
\end{document}